\documentclass[12pt,aps,prd,showpacs,amsmath,amssymb]{revtex4}
\input epsf
\allowdisplaybreaks
\begin{document}
\title{QCD Sum Rule Analysis of Semileptonic $B_{s1}$, $B^{*}_{s2}$, $B^{*}_{s0}$, and $B'_{s1}$ Decays in HQET}
\author{Long-Fei Gan}
 \email{lfgan@nudt.edu.cn}
\author{Ming-Qiu Huang}
\affiliation{Department of Physics, College of Science, National University of Defense Technology, Changsha, Hunan 410073, P. R. China}
\date{\today}
\begin{abstract}
We present an analysis of semileptonic decays of orbitally, $P$-wave excited $B_{s}$ meson states $B^{**}_{s}$, including the newly found narrow $B_{s1}(5830)$ and $B^{*}_{s2}(5840)$ states, into low lying $D_{s}$ mesons ($D_{s}(1968)$, $D^{*}_{s}(2112)$, $D_{sJ}(2317)$, $D_{sJ}(2460)$) within the framework of heavy quark effective theory. The relevant universal form factors are estimated using QCD sum rules at the leading-order of the heavy quark expansion. The decay widths are predicted and the branching ratios are estimated.
\end{abstract}
 \maketitle

\section{Introduction}\label{sec1}
Recently two orbitally excited narrow $B_{s}$ mesons were reported to be observed by the CDF collaboration with masses $m_{B_{s1}}=5829.4 \pm 0.7 \mbox{MeV}$ and $m_{B^{*}_{s2}}=5839.6 \pm 0.7 \mbox{MeV}$ \cite{CDF08}. The D0 collaboration confirmed the $B^{*}_{s2}$ state with mass $m_{B^{*}_{s2}}=5839.6 \pm 1.1 (\text{stat.}) \pm 0.7 (\text{syst.}) \mbox{MeV}$ subsequently, but their searching for a $B_{s1}$ signal gave inconclusive results with the available data set \cite{D008}.  According to heavy quark effective theory (HQET) \cite{Neu94}, the heavy quark spin decouples from the light degrees of freedom in the heavy quark $m_{Q} \rightarrow \infty$ limit. The heavy-light mesons can be categorized according to the total angular momentum of the light degree of freedom $j_{q}$. Combining $j_{q}$ with the spin of the heavy quark yields a doublet of heavy-light meson states $j=j_{q}\pm1/2$. For $S$-wave ground-states with $j_{q}=1/2$ ($L=0$), two states of negative parity form the $H$ doublet $(0^{-}, 1^{-})$. For $P$-wave excited states with $j_{q}=1/2$ or $j_{q}=3/2$ ($L=1$), four states of positive parity form the $S$ doublet $(0^{+}, 1^{+})$ and the $T$ doublet $(1^{+}, 2^{+})$. The two observed excited $B_{s}$ states above are considered as members of the $T$ doublet in the $\bar{b}s$ system \cite{CDF08}.

The four $P$-wave excited $\bar{b}s$ states are often referred to as $B^{**}_{s}$. If kinematically allowed, they are expected to decay into $B^{*}K$, $BK$ or both. The $S$ doublet of $\bar{b}s$ system, ($B^{*}_{s0}$, $B'_{s1}$), are supposed to be broad because they decay through an $S$-wave transition while the $T$ doublet, ($B_{s1}$, $B^{*}_{s2}$), are narrow because they decay through a $D$-wave transition \cite{ZZ08}. This is the case of $B_{s1}$ and $B^{*}_{s2}$ because their masses lie above the thresholds of $B^{*}K$ and $BK$ and they can decay dominantly through these strong channels as is expected. But we don't know the masses of $B^{*}_{s0}$ and $B'_{s1}$ experimentally. In fact, they are speculated to lie below the thresholds of $B^{*}K$ and $BK$ \cite{BEH03}. Therefore, the channels $(B^{*}_{s0}, B'_{s1})\rightarrow (B^{*}, B)+K$ will be kinematically blocked and the isospin violating strong decays and electromagnetic decays will be their dominant decay modes. If so, they should be very narrow as well. This is quite similar to the situation of similarly excited $D_{s}$ mesons \cite{Swa06}.

Lots of work have been done to study the properties of these states and their decays. Their masses had been predicted theoretically in Refs. \cite{BEH03,VVF08,EHQ93} through different methods. The strong decays were investigated in Refs. \cite{LCL09,ZZ08} using different models. Motivated by the discovery of $B_{s1}(5830)$ and $B^{*}_{s2}(5840)$, we investigate semileptonic decays of $B^{**}_{s}$ into the $H$ ($D_{s}(1968)$, $D^{*}_{s}(2112)$) and $S$ ($D_{sJ}(2317)$, $D_{sJ}(2460)$) doublets of $\bar{c}s$ mesons using QCD sum rules in the framework of HQET in this work.

The QCD sum rule method \cite{Shi79}, in combination with HQET, has been widely used to explore the properties and decays of heavy-light mesons and has been proved to be a powerful theoretical apparatus for studying heavy hadrons. In the past twenty years, study of semileptonic decay of the ground-state $B$ and $B_{s}$ mesons by this method and quark models has been extensively explored \cite{LLSW97,Neu92,EFG99,Dea98,Hua99,Col00,Gan09,Hua04,AAO06}. In our previous works, we have investigated the semileptonic decays of ground-state $B$ mesons into some highly excited charmed meson doublets through this method \cite{Gan09}. Based on the general method of parametrizing the hadron matrix elements as proposed in Ref. \cite{Fal92} and the general interpolating currents for heavy-light mesons in HQET \cite{Dai97}, it can be easily extended to the case of the semileptonic decays of excited $B$ and $B_{s}$ mesons. To date, no work about the semileptonic decays of $B^{**}_{s}$ has been done with the same method.

The remainder of this paper is organized as follows. After an introduction, we derive the formulas of weak current matrix elements and decay rates in HQET in Sec. \ref{sec2}. In Sec. \ref{sec3} we deduce the three-point sum rules for the relevant universal form factors. Finally, section \ref{sec4} is devoted to numerical results and discussions.

\section{Decay matrix elements and decay widths} \label{sec2}

The semileptonic decay rate of a $B$ meson transition into a charmed meson is determined by the corresponding matrix elements of the weak axial-vector and vector currents ($V^{\mu} = \overline{c}\gamma^{\mu}b$ and $A^{\mu} = \overline{c}\gamma^{\mu}\gamma_{5}b$) between them. The parametrizations of these hadronic matrix elements are important because it is not easy to calculated them from first principles. They are often parametrized in terms of form factors which describe the momentum distribution among composites of a hadron. In HQET, the number of form factors is reduced dramatically and this brings about much convenience. Especially at the leading order of the heavy quark expansion, the transition between certain low-lying heavy quark doublets can be parametrized by only one Isgur-Wise form factor for each process. This can be done in different methods. In Ref. \cite{LCLZ09}, they finished the parameterizations in CQM model. Here we make use of the trace formalism given in Ref. \cite{Fal92}.

\subsection{$B_{s1}(B^{*}_{s2})\rightarrow D_{s}(D^{*}_{s})\ell\overline{\nu}$ and $B_{s1}(B^{*}_{s2})\rightarrow D_{s0}(D'_{s1})\ell\overline{\nu}$ decays} \label{subsec1}
The heavy-light meson doublets can be expressed conveniently as effective operators. For the decays $B_{s1}(B^{*}_{s2})\rightarrow D_{s}(D^{*}_{s})\ell\overline{\nu}$ and $B_{s1}(B^{*}_{s2})\rightarrow D_{s0}(D'_{s1})\ell\overline{\nu}$, one needs to introduce six field operators to represent the six heavy-light mesons that appear in these processes. The fields $P^{\nu}$ and $P^{*\mu\nu}$ that annihilate members of the initial $T$ doublet with four-velocity $v$ are in the representation
\begin{equation}\label{operator1}
 T^{\mu}_{v}=\frac{1+\rlap/v}{2}\{P^{*\mu\nu}\gamma_{\nu}-\sqrt{\frac{3}{2}}P^{\nu}\gamma_{5}[g^{\mu}_{\nu}-\frac{1}{3}\gamma_{\nu}(\gamma^{\mu}-v^{\mu})]\},
\end{equation}
 where $\rlap/v = v\cdot\gamma$. The fields $P$, $P^{*}_{\mu}$, $P_{0}$, and $P'^{*}_{1\mu}$ that annihilate members of the final $H$ and $S$ doublets with four-velocity $v$ are, in form,
\begin{equation}\label{operator2}
 H_{v}=\frac{1+\rlap/v}{2}[P^{*}_{\mu}\gamma^{\mu}-P\gamma_{5}]
\end{equation}
and
\begin{equation}\label{operator3}
 S_{v}=\frac{1+\rlap/v}{2}[P'^{*}_{1\mu}\gamma^{\mu}\gamma_{5}+P_{0}].
\end{equation}
At the leading order of the heavy quark expansion, the hadronic matrix elements of the weak current between states annihilated by fields in $H_{v'}$ and $T_{v}$ are
\begin{equation}\label{trace1}
 \bar{h}^{(c)}_{v'}\Gamma h^{(b)}_{v}=\xi(y)Tr\{v'_{\alpha}\overline{H}_{v'}\Gamma T^{\alpha}_{v}\},
\end{equation}
where $h^{(Q)}_{v}$ is the heavy quark field in HQET, $\overline{H}_{v'}=\gamma_{0}H_{v'}^{\dag}\gamma_{0}$, and $\xi(y)$ is a universal Isgur-Wise function of the product of velocities $y(=v\cdot v')$. For hadronic elements between states annihilated by fields in $S_{v'}$ and $T_{v}$, they are similarly written as
\begin{equation}\label{trace2}
 \bar{h}^{(c)}_{v'}\Gamma h^{(b)}_{v}=\zeta(y)Tr\{v'_{\alpha}\overline{S}_{v'}\Gamma T^{\alpha}_{v}\}.
\end{equation}
Here we should notice that each side of Eq. (\ref{trace1}) and Eq. (\ref{trace2}) is understood to be inserted between corresponding initial $B_{s}$ and final $D_{s}$ states. The hadronic matrix elements of $B_{s1}(B^{*}_{s2})\rightarrow D_{s}(D^{*}_{s})\ell\overline{\nu}$ can be calculated straightforwardly from the trace formalism (\ref{trace1}) and are given as follows:
\begin{align}\label{matrix1}
\frac{\langle D_{s}(v')|(V-A)^{\mu}|B_{s1}(v,\epsilon)\rangle}{\sqrt{m_{B_{s1}}m_{D_{s}}}} = &
-\sqrt{\frac{1}{6}}\xi(y)
\epsilon_{\beta}[(y^{2}-1)g^{\beta\mu}-(y-2)v^{\mu}v'^{\beta}+3v'^{\mu}v'^{\beta}\nonumber\\&+i(y+1)v_{\rho}v'_{\tau}\varepsilon^{\beta\mu\rho\tau}],
\\\label{matrix2}
\frac{\langle D^{*}_{s}(v',\epsilon')|(V-A)^{\mu}|B_{s1}(v,\epsilon)\rangle}{\sqrt{m_{B_{s1}}m_{D^{*}_{s}}}} = & \sqrt{\frac{1}{6}}
\times\xi(y)\epsilon'^{*}_{\sigma}\epsilon_{\beta}[(y+1)(2g^{\sigma\mu}v'^{\beta}+g^{\beta\mu}v^{\sigma})-3v^{\sigma}
v'^{\mu}v'^{\beta}\nonumber\\&-(y+1)(v^{\mu}-v'^{\mu})g^{\sigma\beta}-i(y-1)
\varepsilon^{\beta\sigma\mu\tau}(v_{\tau}+v'_{\tau})\nonumber\\&+2iv'^{\mu}v_{\rho}v'_{\tau}\varepsilon_{\beta\sigma\rho\tau}
 -iv'^{\beta}v_{\rho}v'_{\tau}\varepsilon^{\sigma\mu\rho\tau}],
\\\label{matrix3}
\frac{\langle D_{s}(v')|(V-A)^{\mu}|B^{*}_{s2}(v,\epsilon)\rangle}{\sqrt{m_{B^{*}_{s2}}m_{D_{s}}}} = &
\xi(y)v'^{\alpha}\epsilon_{\alpha\beta}[(y+1) g^{\beta\mu}-v'^{\beta}v^{\mu}-iv_{\rho}v'_{\tau}\varepsilon^{\beta\mu\rho\tau}],
\\\label{matrix4}
\frac{\langle D^{*}_{s}(v',\epsilon')|(V-A)^{\mu}|B^{*}_{s2}(v,\epsilon)\rangle}{\sqrt{m_{B^{*}_{s2}}m_{D^{*}_{s}}}} = &
\xi(y)\epsilon'^{*}_{\sigma}\epsilon_{\alpha\beta}[(v^{\mu}+v'^{\mu})g^{\sigma\beta}-v^{\sigma}g^{\mu\beta}-v'^{\beta}
g^{\mu\sigma}\nonumber\\&-iv'^{\alpha}\varepsilon^{\beta\sigma\mu\tau}(v_{\tau}+v'_{\tau})].
\end{align}
For the decays $B_{s1}(B^{*}_{s2})\rightarrow D_{s0}(D'_{s1})\ell\overline{\nu}$, the corresponding hadronic matrix elements are calculated from Eq. (\ref{trace2}) as follows:
\begin{align}
\label{matrix5}
\frac{\langle D_{s0}(v')|(V-A)^{\mu}|B_{s1}(v,\epsilon)\rangle}{\sqrt{m_{B_{s1}}m_{D_{s0}}}} = &
\sqrt{\frac{1}{6}}\zeta(y)
\epsilon_{\beta}[(y^{2}-1)g^{\beta\mu}-(y+2)v^{\mu}v'^{\beta}+3v'^{\mu}v'^{\beta}\nonumber\\&-i(y-1)v_{\rho}v'_{\tau}\varepsilon^{\beta\mu\rho\tau}],
\\\label{matrix6}
\frac{\langle D'_{s1}(v',\epsilon')|(V-A)^{\mu}|B_{s1}(v,\epsilon)\rangle}{\sqrt{m_{B_{s1}}m_{D'_{s1}}}} = &
\sqrt{\frac{1}{6}}\zeta(y)\epsilon'^{*}_{\sigma}\epsilon_{\beta}[(y-1)(2g^{\sigma\mu}v'^{\beta}-g^{\beta\mu}v^{\sigma})-3v^{\sigma}
v'^{\mu}v'^{\beta}\nonumber\\&+(y-1)(v^{\mu}+v'^{\mu})g^{\sigma\beta}+i(y+1)
\varepsilon^{\beta\sigma\mu\tau}(v_{\tau}-v'_{\tau})\nonumber\\&+2iv'^{\mu}v_{\rho}v'_{\tau}\varepsilon^{\beta\sigma\rho\tau}
-iv'^{\beta}v_{\rho}v'_{\tau}\varepsilon^{\sigma\mu\rho\tau}],
\\\label{matrix7}
\frac{\langle D_{s0}(v')|(V-A)^{\mu}|B^{*}_{s2}(v,\epsilon)\rangle}{\sqrt{m_{B^{*}_{s2}}m_{D_{s0}}}} = &
\zeta(y)v'^{\alpha}\epsilon_{\alpha\beta}[(1-y)g^{\beta\mu}+v_{\mu}v'_{\beta}+iv_{\rho}v'_{\tau}\varepsilon^{\beta\mu\rho\tau}],
\\\label{matrix8}
\frac{\langle D'_{s1}(v',\epsilon')|(V-A)^{\mu}|B^{*}_{s2}(v,\epsilon)\rangle}{\sqrt{m_{B^{*}_{s2}}m_{D'_{s1}}}} = &
\zeta(y)v'^{\alpha}\epsilon'^{*}_{\sigma}\epsilon_{\alpha\beta}[g^{\beta\mu}v^{\sigma}-
g^{\sigma\mu}v'^{\beta}-(v^{\mu}-v'^{\mu})g^{\beta\sigma}\nonumber\\&-i\varepsilon^{\beta\sigma\mu\tau}(v_{\tau}-v'_{\tau})].
\end{align}
In these matrix elements, $\epsilon_{\alpha}$ ($\epsilon'_{\alpha}$) is the polarization vector of the initial (final) vector meson and $\epsilon_{\alpha\beta}$ is the polarization tensor of the initial tensor meson. $v$ is the velocity of the initial meson and $v'$ is the velocity of the final meson in each process. $(V-A)^{\mu}=\overline{c}\gamma^{\mu}(1-\gamma_{5})b$ is the weak current. Using the matrix elements (\ref{matrix1})-(\ref{matrix8}), we derive the differential decay widths for these processes in terms of the Isgur-Wise functions as follows:
\begin{align}\label{rate1}
\frac{d\Gamma}{dy}(B_{s1}\rightarrow D_{s}\ell\overline{\nu})= &
\frac{G^{2}_{F}|V_{cb}|^{2}m^{2}_{B_{s1}}m^{3}_{D_{s}}}{216\pi^{3}}|\xi(y)|^{2}(y-1)^{\frac{3}{2}}
(y+1)^{\frac{5}{2}}[(1+r_{1}^{2})(2y-1)\nonumber\\&-2r_{1}(y^{2}-y+1)],
\\\label{rate2}
\frac{d\Gamma}{dy}(B_{s1}\rightarrow D^{*}_{s}\ell\overline{\nu})= &
\frac{G^{2}_{F}|V_{cb}|^{2}m^{2}_{B_{s1}}m^{3}_{D^{*}_{s}}}{216\pi^{3}}|\xi(y)|^{2}(y-1)^{\frac{3}{2}}
(y+1)^{\frac{5}{2}}[(1+r_{2}^{2})(7y+1)\nonumber\\&-2r_{2}(5y^{2}+y+2)],
\\\label{rate3}
\frac{d\Gamma}{dy}(B_{s1}\rightarrow D_{s0}\ell\overline{\nu})= &
\frac{G^{2}_{F}|V_{cb}|^{2}m^{2}_{B_{s1}}m^{3}_{D_{s0}}}{216\pi^{3}}|\zeta(y)|^{2}(y-1)^{\frac{5}{2}}
(y+1)^{\frac{3}{2}}[(1+r_{3}^{2})(2y+1)\nonumber\\&-2r_{3}(y^{2}+y+1)],
\\\label{rate4}
\frac{d\Gamma}{dy}(B_{s1}\rightarrow D'_{s1}\ell\overline{\nu})= &
\frac{G^{2}_{F}|V_{cb}|^{2}m^{2}_{B_{s1}}m^{3}_{D'_{s1}}}{216\pi^{3}}|\zeta(y)|^{2}(y-1)^{\frac{5}{2}}
(y+1)^{\frac{3}{2}}[(1+r_{4}^{2})(7y-1)\nonumber\\&-2r_{4}(5y^{2}-y+2)],
\\\label{rate5}
\frac{d\Gamma}{dy}(B^{*}_{s2}\rightarrow D_{s}\ell\overline{\nu})= &
\frac{G^{2}_{F}|V_{cb}|^{2}m^{2}_{B^{*}_{s2}}m^{3}_{D_{s}}}{360\pi^{3}}|\xi(y)|^{2}(y-1)^{\frac{3}{2}}
(y+1)^{\frac{5}{2}}[(1+r_{5}^{2})(4y+1)\nonumber\\&-2r_{5}(3y^{2}+y+1)],
\\\label{rate6}
\frac{d\Gamma}{dy}(B^{*}_{s2}\rightarrow D^{*}_{s}\ell\overline{\nu})= &
\frac{G^{2}_{F}|V_{cb}|^{2}m^{2}_{B^{*}_{s2}}m^{3}_{D^{*}_{s}}}{360\pi^{3}}|\xi(y)|^{2}(y-1)^{\frac{3}{2}}
(y+1)^{\frac{5}{2}}[(1+r_{6}^{2})(11y-1)\nonumber\\&-2r_{6}(7y^{2}-y+4)],
\\\label{rate7}
\frac{d\Gamma}{dy}(B^{*}_{s2}\rightarrow D_{s0}\ell\overline{\nu})= &
\frac{G^{2}_{F}|V_{cb}|^{2}m^{2}_{B^{*}_{s2}}m^{3}_{D_{s0}}}{360\pi^{3}}|\zeta(y)|^{2}(y-1)^{\frac{5}{2}}
(y+1)^{\frac{3}{2}}[(1+r_{7}^{2})(4y-1)\nonumber\\&-2r_{7}(3y^{2}-y+1)],
\\\label{rate8}
\frac{d\Gamma}{dy}(B^{*}_{s2}\rightarrow D'_{s1}\ell\overline{\nu})= &
\frac{G^{2}_{F}|V_{cb}|^{2}m^{2}_{B^{*}_{s2}}m^{3}_{D'_{s1}}}{360\pi^{3}}|\zeta(y)|^{2}(y-1)^{\frac{5}{2}}
(y+1)^{\frac{3}{2}}[(1+r_{8}^{2})(11y+1)\nonumber\\&-2r_{8}(7y^{2}+y+4)],
\end{align}
where $r_{i}$ ($i=1, 2, \cdots, 8$) is the ratio between the mass of the final $D_{s}$ meson and that of the initial $B^{**}_{s}$ meson in each process. For example, $r_{1}=\frac{M_{D_{s}}}{M_{B_{s1}}}$. The only unknown factors in these equations (\ref{rate1})-(\ref{rate8}) are $\xi(y)$ and $\zeta(y)$, which need to be determined by nonperturbative methods.
\subsection{$B_{s0}(B'_{s1})\rightarrow D_{s}(D^{*}_{s})\ell\overline{\nu}$ and $B_{s0}(B'_{s1})\rightarrow D_{s0}(D'_{s1})\ell\overline{\nu}$ decays} \label{subsec2}
The hadronic matrix elements and differential decay widths of $B_{s0}(B'_{s1})\rightarrow D_{s}(D^{*}_{s})\ell\overline{\nu}$ and $B_{s0}(B'_{s1})\rightarrow D_{s0}(D'_{s1})\ell\overline{\nu}$ can be calculated by repeating the procedure in the previous subsection. The only change is that the initial doublet is now the $S$ doublet in stead of the $T$ doublet. So, at the leading order of the heavy quark expansion, the hadronic matrix elements of the weak current between states destroyed by fields in $H_{v'}$ and $S_{v}$ are
\begin{equation}\label{trace3}
 \bar{h}^{(c)}_{v'}\Gamma h^{(b)}_{v}=\chi(y)Tr\{\overline{H}_{v'}\Gamma S_{v}\},
\end{equation}
while those between states annihilated by fields in $S_{v'}$ and $S_{v}$ are
\begin{equation}\label{trace4}
 \bar{h}^{(c)}_{v'}\Gamma h^{(b)}_{v}=\kappa(y)Tr\{\overline{S}_{v'}\Gamma S_{v}\}.
\end{equation}
Using formula (\ref{trace3}) and (\ref{trace4}), one can easily derive the hadronic matrix elements of the decays $B_{s0}(B'_{s1})\rightarrow D_{s}(D^{*}_{s})\ell\overline{\nu}$ and $B_{s0}(B'_{s1})\rightarrow D_{s0}(D'_{s1})\ell\overline{\nu}$ as follows:
\begin{align}\label{matrix17}
\frac{\langle D_{s}(v')|(V-A)^{\mu}|B^{*}_{s0}(v)\rangle}{\sqrt{m_{B^{*}_{s0}}m_{D_{s}}}} = & \chi(y)(v^{\mu}-v'^{\mu}),
\\\label{matrix18}
\frac{\langle D^{*}_{s}(v',\epsilon')|(V-A)^{\mu}|B^{*}_{s0}(v)\rangle}{\sqrt{m_{B^{*}_{s0}}m_{D^{*}_{s}}}} = &
\chi(y)\epsilon'^{*}_{\sigma}[-(y-1)g^{\sigma\mu}+v'^{\mu}v^{\sigma}+iv_{\rho}v'_{\tau}\varepsilon^{\sigma\mu\rho\tau}],
\\\label{matrix19}
\frac{\langle D_{s}(v')|(V-A)^{\mu}|B'_{s1}(v,\epsilon)\rangle}{\sqrt{m_{B'_{s1}}m_{D_{s}}}} = &
\chi(y)\epsilon_{\alpha}[(y-1)g^{\alpha\mu}-v^{\mu}v'^{\alpha}-iv_{\rho}v'_{\tau}\varepsilon^{\alpha\mu\rho\tau}],
\\\label{matrix20}
\frac{\langle D^{*}_{s}(v',\epsilon')|(V-A)^{\mu}|B'_{s1}(v,\epsilon)\rangle}{\sqrt{m_{B'_{s1}}m_{D^{*}_{s}}}} = &
\chi(y)\epsilon'^{*}_{\sigma}\epsilon_{\alpha}[g^{\alpha\sigma}(v^{\mu}-v'^{\mu})+g^{\sigma\mu}v'^{\alpha}-g^{\alpha\mu}v^{\sigma}\nonumber\\&
-i\varepsilon^{\sigma\alpha\mu\tau}(v_{\tau}-v'_{\tau})],
\\\label{matrix21}
\frac{\langle D_{s0}(v')|(V-A)^{\mu}|B^{*}_{s0}(v)\rangle}{\sqrt{m_{B^{*}_{s0}}m_{D_{s0}}}} = & \kappa(y)(v'^{\mu}+v^{\mu}),
\\\label{matrix22}
\frac{\langle D'_{s1}(v',\epsilon')|(V-A)^{\mu}|B^{*}_{s0}(v)\rangle}{\sqrt{m_{B^{*}_{s0}}m_{D'_{s1}}}} = & \kappa(y)\epsilon'^{*}_{\sigma}[-(y+1)g^{\sigma\mu}+v'^{\mu}v^{\sigma}+iv_{\rho}v'_{\tau}\varepsilon^{\sigma\mu\rho\tau}],
\\\label{matrix23}
\frac{\langle D_{s0}(v',\epsilon')|(V-A)^{\mu}|B'_{s1}(v,\epsilon)\rangle}{\sqrt{m_{B'_{s1}}m_{D_{s0}}}} = & \kappa(y)\epsilon_{\alpha}[-(y+1)g^{\alpha\mu}+v^{\mu}v'^{\alpha}+iv_{\rho}v'_{\tau}\varepsilon^{\alpha\mu\rho\tau}],
\\\label{matrix24}
\frac{\langle D'_{s1}(v',\epsilon')|(V-A)^{\mu}|B'_{s1}(v,\epsilon)\rangle}{\sqrt{m_{B'_{s1}}m_{D'_{s1}}}} = &
-\kappa(y)\epsilon'^{*}_{\sigma}\epsilon_{\alpha}[g^{\alpha\sigma}(v^{\mu}+v'^{\mu})-g^{\sigma\mu}v'^{\alpha}-g^{\alpha\mu}v^{\sigma}\nonumber\\&
+i\varepsilon^{\sigma\alpha\mu\tau}(v_{\tau}+v'_{\tau})],
\end{align}
The differential decay widths of these semileptonic processes are readily calculated from these matrix elements. They are given in terms of $\chi(y)$ and $\kappa(y)$ as follows:
\begin{align}\label{rate9}
\frac{d\Gamma}{dy}(B^{*}_{s0}\rightarrow D_{s}\ell\overline{\nu})= &
\frac{G^{2}_{F}|V_{cb}|^{2}m^{2}_{B^{*}_{s0}}m^{3}_{D_{s}}}{48\pi^{3}}|\chi(y)|^{2}(y-1)^{\frac{3}{2}}
(y+1)^{\frac{3}{2}}(r_{9}^{2}-1)^{2},
\\\label{rate10}
\frac{d\Gamma}{dy}(B^{*}_{s0}\rightarrow D^{*}_{s}\ell\overline{\nu})= &
\frac{G^{2}_{F}|V_{cb}|^{2}m^{2}_{B^{*}_{s0}}m^{3}_{D^{*}_{s}}}{48\pi^{3}}|\chi(y)|^{2}(y-1)^{\frac{3}{2}}
(y+1)^{\frac{1}{2}}[(1+r_{10}^{2})(5y-1)\nonumber\\&-2r_{10}(4y^{2}-y+1)],
\\\label{rate11}
\frac{d\Gamma}{dy}(B^{*}_{s0}\rightarrow D_{s0}\ell\overline{\nu})= &
\frac{G^{2}_{F}|V_{cb}|^{2}m^{2}_{B^{*}_{s0}}m^{3}_{D_{s0}}}{48\pi^{3}}|\kappa(y)|^{2}(y-1)^{\frac{3}{2}}
(y+1)^{\frac{3}{2}}(r_{11}+1)^{2},
\\\label{rate12}
\frac{d\Gamma}{dy}(B^{*}_{s0}\rightarrow D'_{s1}\ell\overline{\nu})= &
\frac{G^{2}_{F}|V_{cb}|^{2}m^{2}_{B^{*}_{s0}}m^{3}_{D'_{s1}}}{48\pi^{3}}|\kappa(y)|^{2}(y-1)^{\frac{1}{2}}
(y+1)^{\frac{3}{2}}[(1+r_{12}^{2})(5y+1)\nonumber\\&-2r_{12}(4y^{2}+y+1)],
\\\label{rate13}
\frac{d\Gamma}{dy}(B'_{s1}\rightarrow D_{s}\ell\overline{\nu})= &
\frac{G^{2}_{F}|V_{cb}|^{2}m^{2}_{B'_{s1}}m^{3}_{D_{s}}}{144\pi^{3}}|\chi(y)|^{2}(y-1)^{\frac{3}{2}}
(y+1)^{\frac{1}{2}}[(1+r_{13}^{2})(5y-1)\nonumber\\&-2r_{13}(4y^{2}-y+1)],
\\\label{rate14}
\frac{d\Gamma}{dy}(B'_{s1}\rightarrow D^{*}_{s}\ell\overline{\nu})= &
\frac{G^{2}_{F}|V_{cb}|^{2}m^{2}_{B'_{s1}}m^{3}_{D^{*}_{s}}}{144\pi^{3}}|\chi(y)|^{2}(y-1)^{\frac{3}{2}}
(y+1)^{\frac{1}{2}}[(1+r_{14}^{2})(13y+1)\nonumber\\&-2r_{14}(8y^{2}+y+5)],
\\\label{rate15}
\frac{d\Gamma}{dy}(B'_{s1}\rightarrow D_{s0}\ell\overline{\nu})= &
\frac{G^{2}_{F}|V_{cb}|^{2}m^{2}_{B'_{s1}}m^{3}_{D_{s0}}}{144\pi^{3}}|\kappa(y)|^{2}(y-1)^{\frac{1}{2}}
(y+1)^{\frac{3}{2}}[(1+r_{15}^{2})(5y+1)\nonumber\\&-2r_{15}(4y^{2}+y+1)],
\\\label{rate16}
\frac{d\Gamma}{dy}(B'_{s1}\rightarrow D'_{s1}\ell\overline{\nu})= &
\frac{G^{2}_{F}|V_{cb}|^{2}m^{2}_{B'_{s1}}m^{3}_{D'_{s1}}}{144\pi^{3}}|\kappa(y)|^{2}(y-1)^{\frac{1}{2}}
(y+1)^{\frac{3}{2}}[(1+r_{16}^{2})(13y-1)\nonumber\\&-2r_{16}(8y^{2}-y+5)].
\end{align}
The definition of $r_{i}$ ($i=9,\cdots, 16$) is the same as $r_{i}$ ($i=1,\cdots, 8$) in the previous subsection. The only unknown factors in these equations, (\ref{rate9})-(\ref{rate16}), now are $\chi(y)$ and $\kappa(y)$ which are also nonpurterbative quantities. In the following section, we will apply the QCD sum rule approach to estimate these Isgur-Wise functions $\xi(y)$, $\zeta(y)$, $\chi(y)$ and $\kappa(y)$.

\section{Form factors from HQET sum rules}\label{sec3}
\subsection{$\xi(y)$ and $\zeta(y)$}\label{subsec3}
In order to apply QCD sum rules to study these heavy mesons, we must choose appropriate interpolating currents to represent these states. Here we adopt the interpolating currents proposed in Ref. \cite{Dai97} based on the study of Bethe-Salpeter equations for heavy mesons in HQET. These currents $J^{\alpha_{1}\cdots\alpha_{j}}_{j,P,j_{\ell}}$ for the states with the quantum numbers $j$, $P$, $j_{l}$ in HQET are proved to have nice properties and satisfy
\begin{equation}\label{property1}
\langle 0|J^{\alpha_{1}\cdots\alpha_{j}}_{j,P,j_{\ell}}(0)|j',P',j'_{\ell}\rangle =f_{Pj_{\ell}}\delta_{jj'}\delta_{PP'}\delta_{j_{\ell}j'_{\ell}}
\eta^{\alpha_{1}\cdots\alpha_{j}}
\end{equation}
and
\begin{equation}\label{property2}
i\langle 0|T[J^{\alpha_{1}\cdots\alpha_{j}}_{j,P,j_{\ell}}(x)J^{\dag \beta_{1}\cdots\beta_{j'}}_{j',P',j'_{\ell}}(0)]|0\rangle =\delta_{jj'}\delta_{PP'}\delta_{j_{\ell}j'_{\ell}}(-1)^{j}\emph{S}g^{\alpha_{1}\beta_{1}}_{t} \cdots g^{\alpha_{j}\beta_{j}}_{t}\int \textsl{d}t \delta(x-vt)\Pi_{P,j_{\ell}}(x),
\end{equation}
in the limit $m_{Q}\rightarrow\infty$, where $\eta^{\alpha_{1}\cdots\alpha_{j}}$ is the polarization tensor for the state with spin $j$, $g^{\alpha\beta}_{t}=g^{\alpha\beta}-v^{\alpha}v^{\beta}$ is the transverse metric tensor, and $\emph{S}$ means to symmetrize the indices and subtract the trace terms separately in the sets ($\alpha_{1}\cdots\alpha_{j}$) and ($\beta_{1}\cdots\beta_{j}$). $f_{Pj_{\ell}}$ is a constant and $\Pi_{P,j_{\ell}}(x)$ is a function of $x$, both of which depend on $P$ (the parity of the meson) and $j_{\ell}$ (the total angular momentum of the light part).

Following the remarks given in Ref. \cite{Gan09}, we take the interpolating currents that create the initial excited $B_{s}$ doublet ($B_{s1}$, $B^{*}_{s2}$) as
\begin{equation}\label{current1}
J^{\dag\alpha}_{1,+,3/2}=(-i)\sqrt{\frac{3}{4}}\bar{h}_{v}\gamma_{5}(D^{\alpha}_{t}-\frac{1}{3}\gamma^{\alpha}_{t}\rlap/D_{t})s,
\end{equation}
\begin{equation}\label{current2}
J^{\dag\alpha\beta}_{2,+,3/2}=\frac{(-i)}{\sqrt{2}}T^{\alpha\beta,\mu\nu}\bar{h}_{v}\gamma_{t\mu}D_{t\nu}s,
\end{equation}
where $D^{\alpha}_{t}=D^{\alpha}-v^{\alpha}(v\cdot D)$ is the transverse component of the covariant derivative with respect to the velocity of the meson. The tensor $T^{\alpha\beta,\mu\nu}$ is used to symmetrize the indices and is given by
\begin{equation}\label{tensor1}
T^{\alpha\beta,\mu\nu}=\frac{1}{2}(g^{\alpha\mu}_{t}g^{\beta\nu}_{t}
+g^{\alpha\nu}_{t}g^{\beta\mu}_{t})-\frac{1}{3}g^{\alpha\beta}_{t}g^{\mu\nu}_{t}.
\end{equation}
For the final charm-strange mesons, the currents correspond to the ground states $D_{s}(1968)$ and $D^{*}_{s}(2112)$ with spin-parity $(0^{-},1^{-})$ are
\begin{equation}\label{current5}
J^{\dag}_{0,-,1/2}=\frac{1}{\sqrt{2}}\bar{h}_{v}\gamma_{5}s,
\end{equation}
\begin{equation}\label{current6}
J^{\dag\alpha}_{1,-,1/2}=\frac{1}{\sqrt{2}}\bar{h}_{v}\gamma^{\alpha}_{t}s,
\end{equation}
and the currents that create the ($D_{sJ}(2307)$, $D_{sJ}(2460)$) doublet with spin-parity $(0^{+},1^{+})$ are
\begin{equation}\label{current3}
J^{\dag}_{0,+,1/2}=\frac{1}{\sqrt{2}}\bar{h}_{v}(-i)\rlap/D_{t}s,
\end{equation}
\begin{equation}\label{current4}
J^{\dag\alpha}_{1,+,1/2}=\frac{1}{\sqrt{2}}\bar{h}_{v}\gamma_{5}\gamma^{\alpha}_{t}(-i)\rlap/D_{t}s.
\end{equation}

With these currents, we can now estimate the Isgur-Wise functions $\xi(y)$ and $\zeta(y)$ using QCD sum rules. We follow the same procedure as Ref. \cite{Neu92} to study the analytical properties of the three-point correlators:
\begin{equation}\label{correlator1}
i^{2}\int d^{4}xd^{4}ze^{i(k^{'}\cdot x-k\cdot
z)}\langle0|T[J_{0,-,1/2}(x)
J^{\mu(v,v^{'})}_{V,A}(0)J^{\alpha\dag}_{1,+,3/2}(z)|0\rangle=\Xi_{1}(\omega,\omega^{'},y)\mathcal
{L}^{\mu\alpha}_{V,A},
\end{equation}
\begin{equation}\label{correlator2}
i^{2}\int d^{4}xd^{4}ze^{i(k^{'}\cdot x-k\cdot
z)}\langle0|T[J_{0,+,1/2}(x)
J^{\mu(v,v^{'})}_{V,A}(0)J^{\alpha\dag}_{1,+,3/2}(z)|0\rangle=\Xi_{2}(\omega,\omega^{'},y)\mathcal
{L'}^{\mu\alpha}_{V,A},
\end{equation}
where $J^{\mu(v,v^{'})}_{V}=h(v^{'})\gamma^{\mu}h(v)$ and $J^{\mu(v,v^{'})}_{A}=h(v^{'})\gamma^{\mu}\gamma_{5}h(v)$. The variables $k$($=P-m_{b}v$) and $k^{'}$($=P'-m_{c}v'$) denote residual ``off-shell" momenta of the initial and final meson states, respectively. For heavy quarks in bound states they are typically of order $\Lambda_{QCD}$ and remain finite in the heavy quark limit. $\Xi_{i}(\omega,\omega^{'},y)$ ($i=1, 2$) are analytic functions in the ``off-shell" energies $\omega=2v \cdot k$ and $\omega'=2v' \cdot k'$ with discontinuities for positive values of these variables. They also depend on the velocity transfer $y=v \cdot v'$, which is fixed in a physical region. $\mathcal {L}^{\mu\alpha}_{V,A}$ and $\mathcal {L'}^{\mu\alpha}_{V,A}$ are Lorentz structures.

Following the standard QCD sum rule procedure the calculations of $\Xi_{i}(\omega,\omega^{'},y)$ ($i=1, 2$) are straightforward. Take (\ref{correlator1}) as an example. First, we saturate Eq. (\ref{correlator1}) with physical intermediate states in HQET and find the hadronic representation of the correlator as follows:
\begin{equation}\label{pheno}
\Xi_{1,hadron}(\omega,\omega^{'},y)=\frac{f_{-,\frac{1}{2}}f_{+,\frac{3}{2}}\xi(y)}
{(2\bar{\Lambda}_{+,\frac{3}{2}}-\omega-i\varepsilon)(2\bar{\Lambda}_{-,\frac{1}{2}}
-\omega^{'}-i\varepsilon)}+ \text{higher resonances},
\end{equation}
where $f_{-,\frac{1}{2}}$ and $f_{+,\frac{3}{2}}$ are the decay constants defined in Eq.(\ref{property1}), $\overline{\Lambda}_{P,j_{l}}$ is the bounding energy of the heavy meson with total parity $P$ and angular momentum of the light part $j_{l}$. Second, the function $\Xi_{1}(\omega,\omega^{'},y)$ can be approximated by a perturbative calculation supplemented by nonperturbative power corrections proportional to the vacuum condensates which are treated as phenomenological parameters. The perturbative contribution can be represented by a double dispersion integral in $\nu$ and $\nu^{'}$ plus possible subtraction terms. So the theoretical expression for the correlator has the form
\begin{equation}\label{theo}
\Xi_{1,theo}(\omega,\omega^{'},y)\simeq\int d\nu d\nu^{'}
\frac{\rho_{1}^{pert}(\nu,\nu^{'},y)}
{(\nu-\omega-i\varepsilon)(\nu^{'}-\omega^{'}-i\varepsilon)}+
\text{subtractions}+\Xi_{1}^{cond}(\omega,\omega^{'},y),
\end{equation}
where $\Xi_{1}^{cond}(\omega,\omega^{'},y)$ is the nonperturbative contribution containing vacuum condensates. The perturbative spectral density and the coefficients of vacuum condensations can be calculated straightforwardly. The other correlator (\ref{correlator2}) can be dealt with in the same way.

Assuming quark-hadron duality, the contributions from higher resonances are usually approximated by the perturbative continuum above a threshold.
Equating the phenomenon and theoretical representations, the contributions of higher resonances in the phenomenon representation (\ref{pheno}) can be eliminated. Following the arguments in Refs. \cite{Neu92,Blo93}, the perturbative and the hadronic spectral densities cannot be locally dual to each other; the necessary way to restore duality is to integrate the spectral densities over the ``off-diagonal" variable $\nu_{-}=\nu-\nu^{'}$, keeping the ``diagonal" variable $\nu_{+}=\frac{\nu+\nu^{'}}{2}$ fixed. It is in $\nu_{+}$ that the
quark-hadron duality is assumed for the integrated spectral densities. The integration region can be expressed in terms of the variables $\nu_{-}$ and $\nu_{+}$ and we choose the triangular region defined by the bounds: $0\leq \nu_{+}\leq \omega_{c}$, $-2\sqrt{\frac{y-1}{y+1}}\nu_{+}\leq \nu_{-}\leq
2\sqrt{\frac{y-1}{y+1}}\nu_{+}$. A double Borel transformation in $\omega$ and $\omega^{'}$ is performed on both sides of the sum rules, in which for simplicity we take the Borel parameters equal
\cite{Neu92,Hua99,Col00}: $T_{1}=T_{2}=2T$. It eliminates the substraction terms in the dispersion integral (\ref{theo}) and improves the convergence of the operator product expansion (OPE) series. Our calculation is confined at the leading order of perturbation. Among the operators in the OPE series, only those with dimension $D \leq  5$ are included. For the condensates of higher dimension ($D > 5$), their values are negligibly small and their contributions are suppressed by the double Borel transformation. So they can be safely omitted. Finally, we obtain the sum rules for the form factors $\xi(y)$ and $\zeta(y)$ as follows:
\begin{align}\label{rule1}
\xi(y)f_{-,1/2}f_{+,3/2}e^{-(\bar{\Lambda}_{-,1/2}+\bar{\Lambda}_{+,3/2})/T}= &
\frac{1}{8\pi^{2}}\frac{1}{(y+1)^{3}}\int^{\omega_{c1}}_{2m_{s}}d\nu_{+}e^{-\frac{\nu_{+}}{T}}
[4\nu^{3}_{+}+3m_{s}(y+1)\nu^{2}_{+} \nonumber\\& -6m^{2}_{s}(y+1)\nu_{+}]
-\frac{\langle\bar{s}s\rangle}{8T}m^{2}_{s}-\frac{\langle g_{s} \bar{s}\sigma\cdot G s\rangle}{24T}
[1-\frac{m_{s}}{4T}\nonumber\\&+\frac{4y+5}{12}\frac{m^{2}_{s}}{T^{2}}]
-\frac{1}{3\times2^{5}}\frac{y+5}{(y+1)^{2}}\langle\frac{\alpha_{s}}{\pi}GG\rangle,
\\\label{rule2}
\zeta(y)f_{+,1/2}f_{+,3/2}e^{-(\bar{\Lambda}_{+,1/2}+\bar{\Lambda}_{+,3/2})/T}= &
\frac{1}{16\pi^{2}}\frac{1}{(y+1)^{3}}\int^{\omega_{c2}}_{2m_{s}}d\nu_{+}e^{-\frac{\nu_{+}}{T}}
[\nu^{4}_{+}-6m_{s}(y+1)\nu^{3}_{+}\nonumber\\&+27m^{2}_{s}(y+1)\nu^{2}_{+}]
+\frac{\langle g_{s} \bar{s}\sigma\cdot G s\rangle}{4}(\frac{1}{3}-\frac{m_{s}}{2T})\nonumber\\&
+\frac{T}{16}\frac{y}{(y+1)^{2}}\langle\frac{\alpha_{s}}{\pi}GG\rangle,
\end{align}

\subsection{$\chi(y)$ and $\kappa(y)$}\label{subsec4}
The derivation of sum rules for $\chi(y)$ and $\kappa(y)$ is just the same. The initial states are now members of the doublet ($B^{*}_{s0}$, $B'_{s1}$) with spin-parity $(0^{+},1^{+})$, so the interpolating currents one should use here are (\ref{current3}) and (\ref{current4}). The final states are the same as those in the previous subsection. Different from the previous subsection, the correlators whose analytical properties we study now are as follows:
\begin{equation}\label{correlator3}
i^{2}\int d^{4}xd^{4}ze^{i(k^{'}\cdot x-k\cdot
z)}\langle0|T[J_{0,-,1/2}(x)
J^{\mu(v,v^{'})}_{V,A}(0)J^{\dag}_{0,+,1/2}(z)|0\rangle=\Xi_{3}(\omega,\omega^{'},y)\mathcal
{L}^{\mu}_{V,A},
\end{equation}
\begin{equation}\label{correlator4}
i^{2}\int d^{4}xd^{4}ze^{i(k^{'}\cdot x-k\cdot
z)}\langle0|T[J_{0,+,1/2}(x)
J^{\mu(v,v^{'})}_{V,A}(0)J^{\dag}_{0,+,1/2}(z)|0\rangle=\Xi_{4}(\omega,\omega^{'},y)\mathcal
{L'}^{\mu}_{V,A},
\end{equation}
By repeating the procedure in subsection \ref{subsec3}, we reach the sum rules for $\chi(y)$ and $\kappa(y)$ in HQET as below:
\begin{align}\label{rule3}
\chi(y)f_{-,1/2}f_{+,1/2}e^{-(\bar{\Lambda}_{-,1/2}+\bar{\Lambda}_{+,1/2})/T}= &
(-\frac{1}{8\pi^{2}})\frac{1}{(y+1)^{2}}\int^{\omega_{c3}}_{2m_{s}}d\nu_{+}e^{-\frac{\nu_{+}}{T}}
[\nu^{3}_{+}-3m_{s}(y+1)\nu^{2}_{+} \nonumber\\& +6m^{2}_{s}(y+1)\nu_{+}]
+\frac{\langle\bar{s}s\rangle}{8}m_{s}[3+m_{s}\frac{y+1}{T}] \nonumber\\& +\frac{\langle g_{s} \bar{s}\sigma\cdot G s\rangle}{24T}
[(2y-1)-m_{s}\frac{y+7}{4T}]\nonumber\\&
-\frac{7}{3\times2^{6}}\frac{y-1}{y+1}\langle\frac{\alpha_{s}}{\pi}GG\rangle,
\\\label{rule4}
\kappa(y)f^{2}_{+,1/2}e^{-2\bar{\Lambda}_{+,1/2}/T}=&
\frac{1}{8\pi^{2}}\frac{1}{(y+1)^{3}}\int^{\omega_{c4}}_{2m_{s}}d\nu_{+}e^{-\frac{\nu_{+}}{T}}
[(y+2)\nu^{4}_{+}\nonumber\\&+3m_{s}y(y+1)\nu^{3}_{+}-3m^{2}_{s}(y+1)(2y+1)\nu^{2}_{+}]
\nonumber\\& -\frac{\langle g_{s} \bar{s}\sigma\cdot G s\rangle}{48}(4y-1)(1-\frac{m_{s}}{T})
\nonumber\\& -\frac{T}{16}\frac{y^{2}+y-2}{(y+1)^{2}}\langle\frac{\alpha_{s}}{\pi}GG\rangle.
\end{align}

\section{Numerical results and discussions}\label{sec4}
We now turn to the numerical evaluation of these sum rules and their phenomenological implications. The input parameters are as follows. The mass of the initial and final heavy mesons are $M_{B_{s1}}=5829.4\mbox{MeV}$, $M_{B'_{s1}}=5765\mbox{MeV}$, $M_{D_{s}}=1968.5\mbox{MeV}$, $M_{D^{*}_{s}}=2112.3\mbox{MeV}$, $M_{D_{s0}}=2317.8\mbox{MeV}$, $M_{D'_{s1}}=2459.6\mbox{MeV}$ \cite{PDG08}, $M_{B^{*}_{s2}}=5839.7\mbox{MeV}$, and $M_{B_{s0}}=5718\mbox{MeV}$ \cite{BEH03}. For the QCD parameters entering the theoretical expressions, we take the standard values: $\label{qcond}\langle\overline{s}s\rangle=-0.8\times(0.24)^{3} \mbox{GeV}^{3}$, $\label{gcond}\langle\alpha_{s}GG\rangle=0.04 \mbox{GeV}^{4}$ and $\label{mcond}m^{2}_{0}=0.8 \mbox{GeV}^{2}$. The mass of the strange quark is $m_{s}=150\mbox{MeV}$. In addition, $V_{cb}=0.04$ and $G_{F}=1.166\times10^{-5}\mbox{GeV}^{-2}$. The cutoff parameter is chosen as $\mu=1 \mbox{GeV}$ \cite{DHLZ03}.

Let's evaluate the sum rule for $\xi(y)$ numerically first. As we can see in Eq.(\ref{rule1}), two decay constants ($f_{+,3/2}$ and $f_{-,1/2}$) and two bounding energies ($\overline{\Lambda}_{+,3/2}$ and $\overline{\Lambda}_{-,1/2}$) appear on the left-hand side as input parameters. In order to obtain information of Isgur-Wise function $\xi(y)$ with less systematic uncertainty, we can remove these parameters by dividing the three-point sum rules with the square roots of relevant two-point sum rules, as many authors did \cite{Neu92,Hua99,Col00}. This can not only reduce the number of input parameters but also improve stabilities of the three-point sum rules. The two-point QCD sum rules we need here are \cite{Hua04,DHLZ03}
\begin{align}\label{conrule1}
f^{2}_{+,3/2}e^{-2\bar{\Lambda}_{+,3/2}/T}= & \frac{1}{64\pi^{2}}\int^{\omega_{2}}_{2m_{s}}d\nu e^{-\frac{\nu}{T}} (\nu^{4}+2m_{s}\nu^{3}-6m^{2}_{s}\nu^{2}-12m^{3}_{s}\nu)\nonumber\\&-\frac{1}{12}m^{2}_{0}\langle \bar{s}s\rangle-\frac{1}{32}\langle\frac{\alpha_{s}}{\pi}GG\rangle T+\frac{1}{8}m^{2}_{s}\langle \bar{s}s\rangle-\frac{m_{s}}{48}\langle\frac{\alpha_{s}}{\pi}GG\rangle,
\\\label{conrule2}
f^{2}_{-,1/2}e^{-2\bar{\Lambda}_{-,1/2}/T}= & \frac{3}{16\pi^{2}}\int^{\omega_{0}}_{2m_{s}}d\nu e^{-\frac{\nu}{T}} (\nu^{2}+2m_{s}\nu-2m^{2}_{s}) -\frac{1}{2}\langle \bar{s}s\rangle(1-\frac{m_{s}}{2T}+\frac{m^{2}_{s}}{2T^{2}})\nonumber\\& +\frac{m^{2}_{0}}{8T^{2}}\langle \bar{s}s\rangle(1-\frac{m_{s}}{3T}+\frac{m^{2}_{s}}{3T^{2}})-\frac{m_{s}}{16T^{2}}\langle\frac{\alpha_{s}}{\pi}GG\rangle(2\gamma_{E}-1-
\text{ln}\frac{T^{2}}{\mu^{2}}).
\end{align}
After the division has been done, we obtain an expression for the $\xi(y)$ as a function of the Borel parameter $T$ and the continuum thresholds $\omega_{c1}$, $\omega_{0}$ and $\omega_{2}$. Imposing usual criteria for the upper and lower bounds of the Borel parameter, we found they have a common sum rule ``window": $0.7\mbox{GeV}<T<1.5\mbox{GeV}$, which overlaps with those of the two-point sum rules (\ref{conrule1}) and (\ref{conrule2}). Notice that the Borel parameter in the sum rule for three-point correlator is twice the Borel parameter in the sum rules for the two-point correlators. In the evaluation we have taken $1.9\mbox{GeV}<\omega_{0}<2.4\mbox{GeV}$ \cite{Hua99,Neu92} and $2.8\mbox{GeV}<\omega_{2}<3.2\mbox{GeV}$ \cite{DHLZ03}. The regions of these continuum thresholds are fixed by analyzing the corresponding two-point sum rules. Following discussions in Refs. \cite{Blo93,Neu92}, the upper limit $\omega_{c1}$ for $\nu_{+}$ in the region
$\frac{1}{2}[(y+1)-\sqrt{y^{2}-1}]\omega_{0}\leqslant\omega_{c1}\leqslant\frac{1}{2}(\omega_{0}+\omega_{2})$ is reasonable. So we can fix $\omega_{c}$ in the region $2.4\mbox{GeV}<\omega_{c1}<2.8\mbox{GeV}$. The results are showed in Fig. 1, in which we fix $\omega_{0}=2.2\mbox{GeV}$ and $\omega_{2}=3.0\mbox{GeV}$.
\begin{figure}
\begin{center}
\begin{tabular}{ccc}
\begin{minipage}{7cm} \epsfxsize=7cm
\centerline{\epsffile{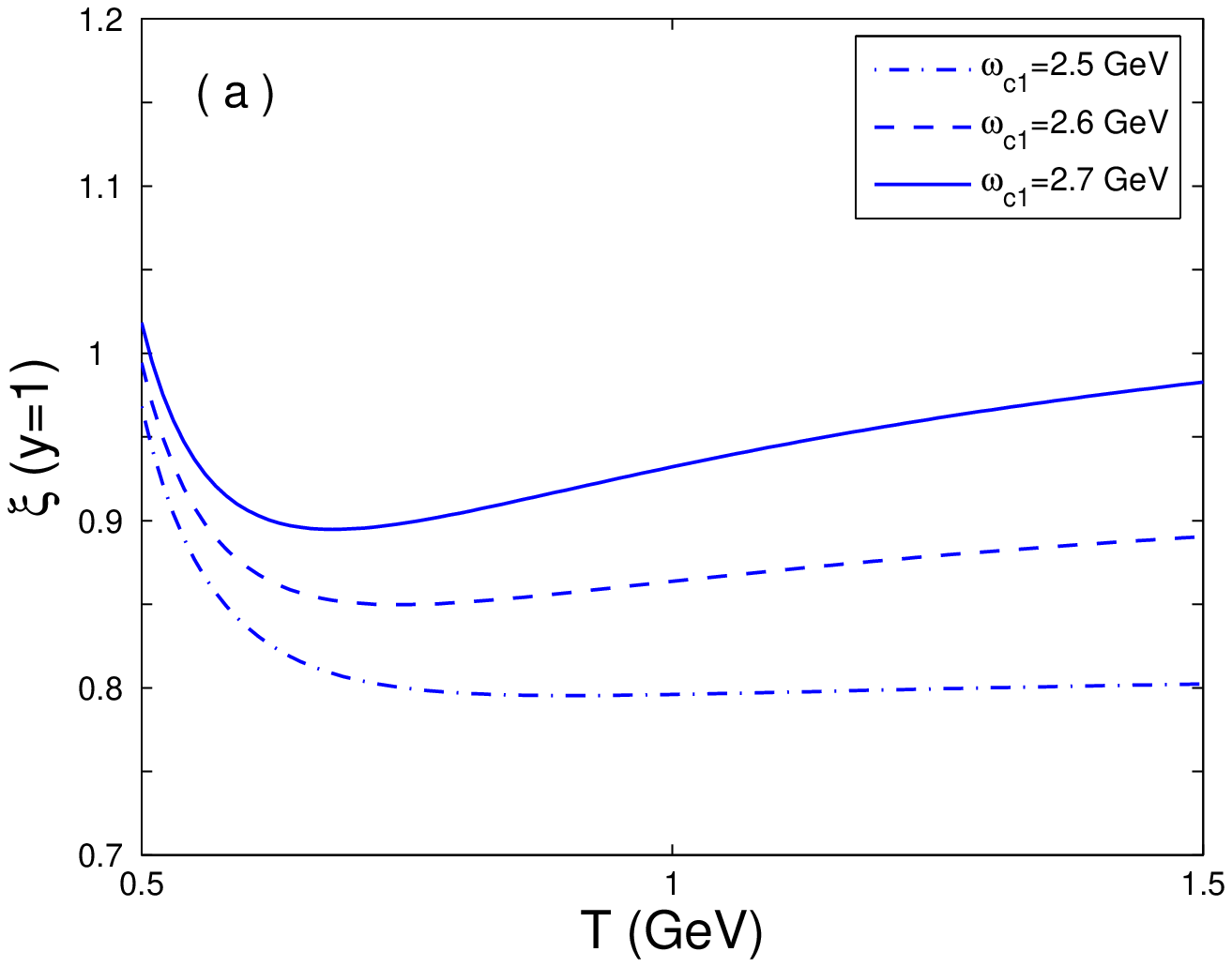}}
\end{minipage}& &
\begin{minipage}{7cm} \epsfxsize=7cm
\centerline{\epsffile{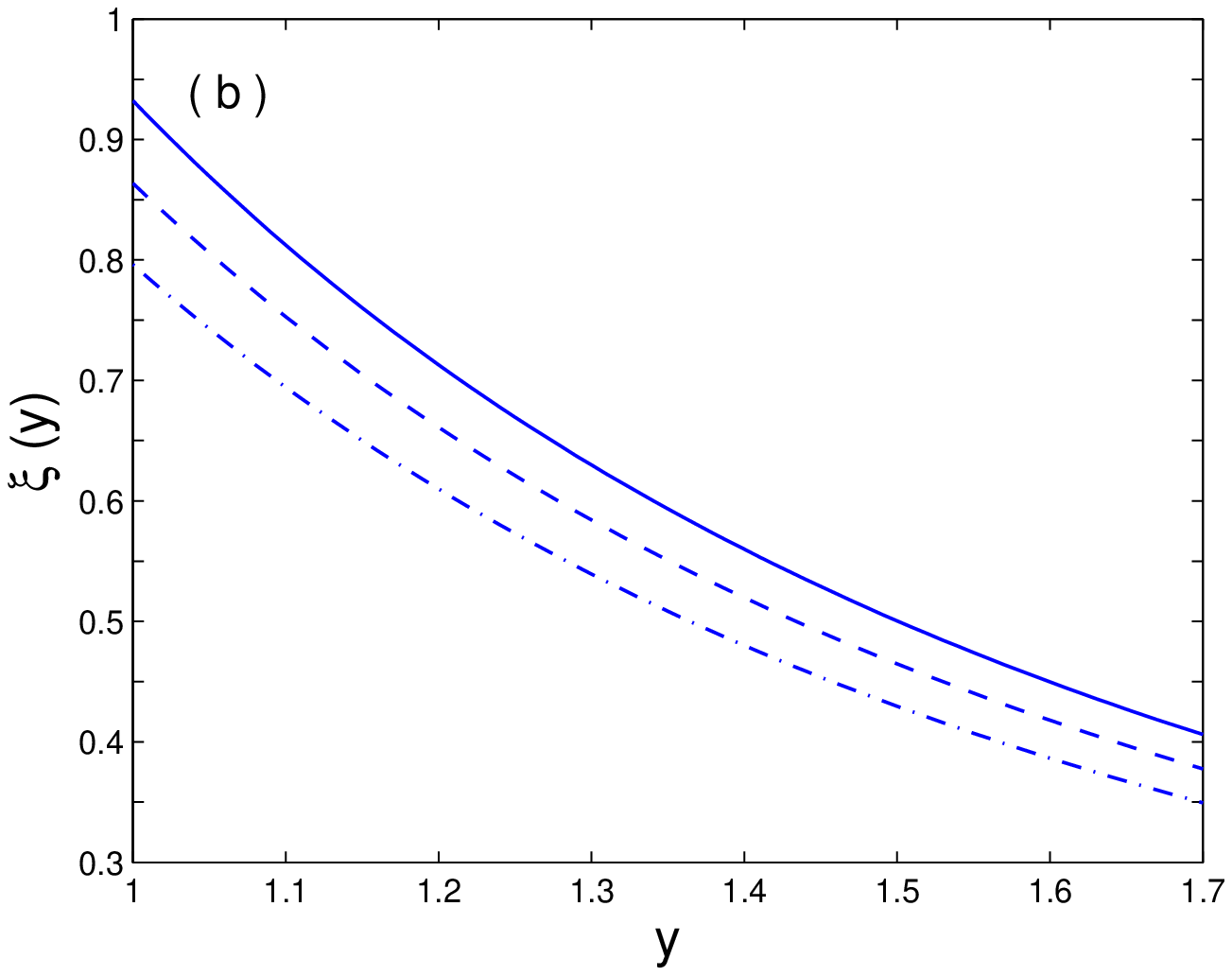}}
\end{minipage}
\end{tabular}
\caption{(a) {\it Dependence of  $\xi(y)$  on Borel parameter $T$ at $y=1$.} (b) {\it Prediction for the Isgur-Wise functions $\xi(y)$ at} $T=1 \mbox{GeV}$.}
\end{center}\label{fig1}
\end{figure}
The resulting curve for $\xi(y)$ can be well fitted by the linear approximation
\begin{equation}\label{linear1}
\xi(y)=\xi(1)[1-\rho^{2}_{\xi}(y-1)],\text{ }
\xi(1)=0.81\pm0.09,\text{ }\rho^{2}_{\xi}=0.83\pm0.06.
\end{equation}
The errors reflect the uncertainty due to $\omega_{c}$ and $T$.

The numerical estimation of $\zeta(y)$ can be done in the same way. The two point-sum rules we need to remove low energy parameters $f_{+,3/2}$, $f_{+,1/2}$, $\overline{\Lambda}_{+,3/2}$ and $\overline{\Lambda}_{+,1/2}$ from Eq. (\ref{rule2}) are Eq. (\ref{conrule1}) and \cite{DHLZ03}
\begin{align}\label{conrule3}
f^{2}_{+,1/2}e^{-2\bar{\Lambda}_{+,1/2}/T}=& \frac{3}{64\pi^{2}}\int^{\omega_{1}}_{2m_{s}}d\nu e^{-\frac{\nu}{T}} (\nu^{4}+2m_{s}\nu^{3}-6m^{2}_{s}\nu^{2}-12m^{3}_{s}\nu)\nonumber\\& -\frac{1}{16}m^{2}_{0}\langle \bar{s}s\rangle(1-\frac{m_{s}}{T}+\frac{4}{3}\frac{m^{2}_{s}}{T^{2}})+\frac{3}{8}m^{2}_{s}\langle \bar{s}s\rangle-\frac{m_{s}}{16}\langle\frac{\alpha_{s}}{\pi}GG\rangle.
\end{align}
The Isgur-Wise function $\zeta(y)$ finally appears as a function of the Borel parameter $T$ and the continuum thresholds $\omega_{c2}$, $\omega_{1}$ and $\omega_{2}$. The region of $\omega_{2}$ is given above while $\omega_{1}$ is fixed in the interval $2.7\mbox{GeV}<\omega_{1}<3.1\mbox{GeV}$ \cite{DHLZ03}. The upper limit $\omega_{c2}$ for $\nu_{+}$ in the region
$\frac{1}{2}[(y+1)-\sqrt{y^{2}-1}]\omega_{1}\leqslant\omega_{c2}\leqslant\frac{1}{2}(\omega_{1}+\omega_{2})$ is reasonable. So it can be fixed in the region $2.7\mbox{GeV}<\omega_{c2}<3.1\mbox{GeV}$. The results are shown in Fig. 2 where we fix $\omega_{1}=2.9\mbox{GeV}$ and $\omega_{2}=3.0\mbox{GeV}$.
\begin{figure}
\begin{center}
\begin{tabular}{ccc}
\begin{minipage}{7cm} \epsfxsize=7cm
\centerline{\epsffile{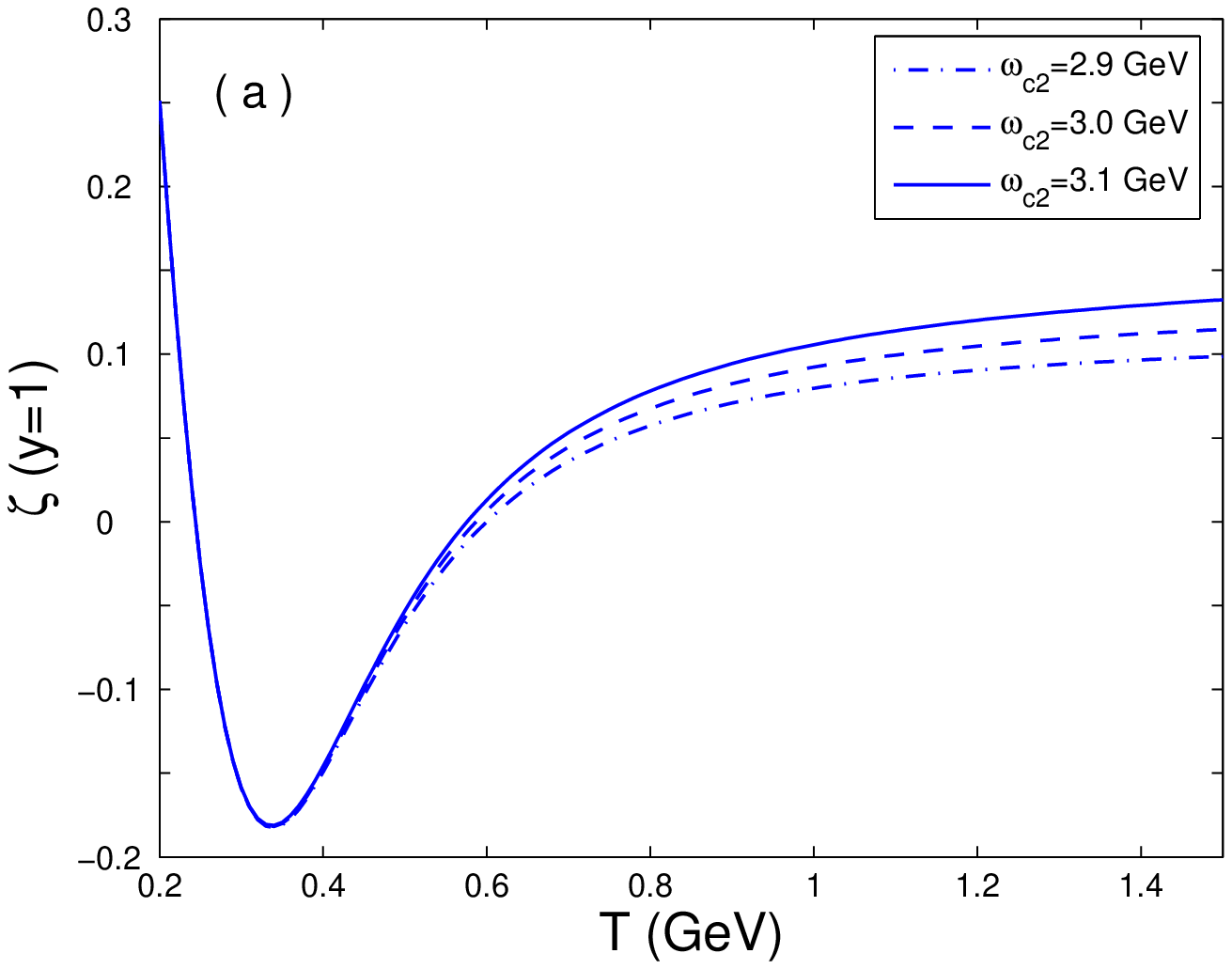}}
\end{minipage}& &
\begin{minipage}{7cm} \epsfxsize=7cm
\centerline{\epsffile{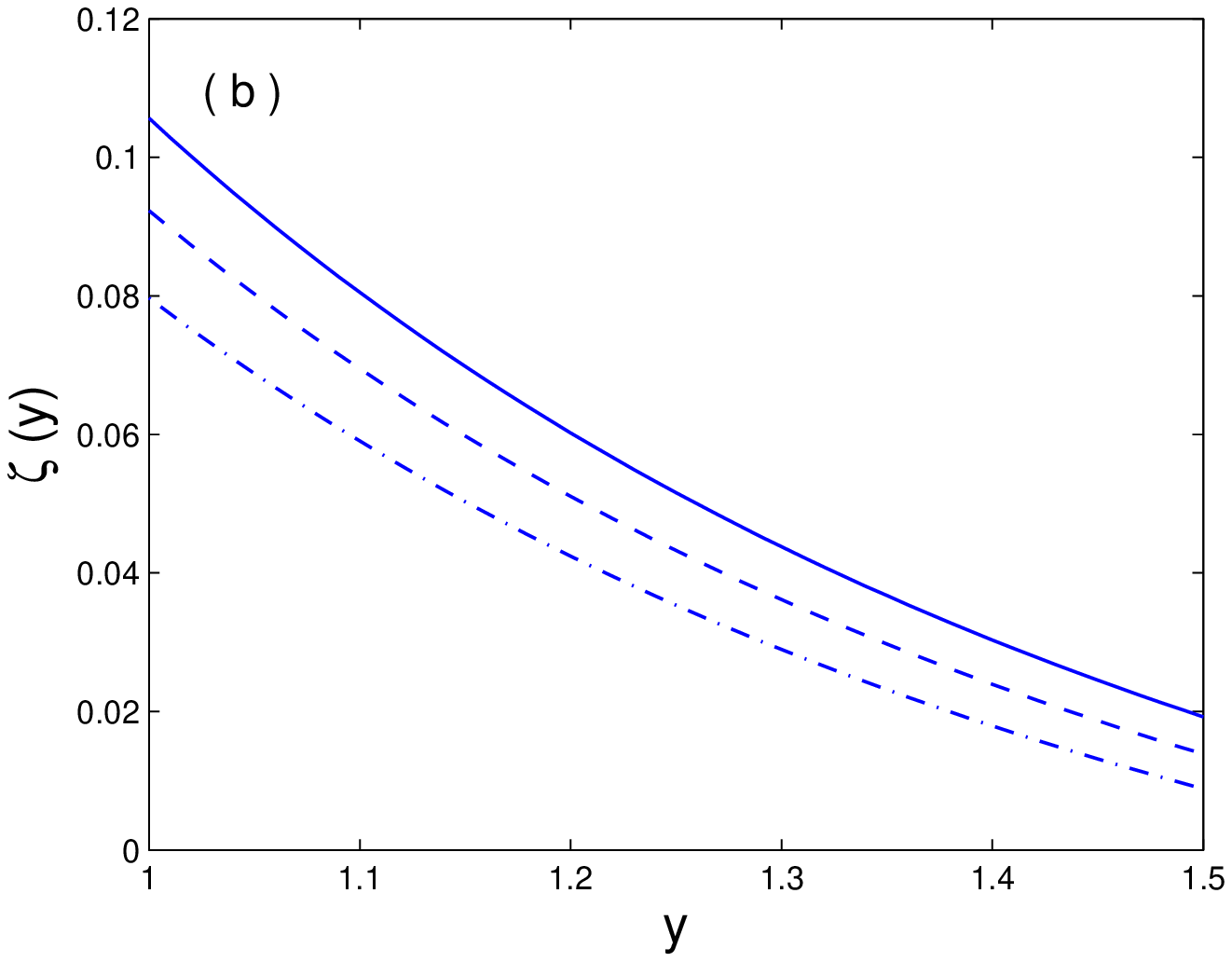}}
\end{minipage}
\end{tabular}
\caption{(a) {\it Dependence of  $\zeta(y)$  on Borel parameter $T$ at $y=1$.} (b) {\it Prediction for the Isgur-Wise functions $\zeta(y)$ at} $T=1 \mbox{GeV}$.}
\end{center}\label{fig2}
\end{figure}
As has been done to $\xi(y)$, the curve of $\zeta(y)$ is also fitted by a linear approximation
\begin{equation}\label{linear2}
\zeta(y)=\zeta(1)[1-\rho^{2}_{\zeta}(y-1)],\text{ }
\zeta(1)=0.085\pm0.010,\text{ }\rho^{2}_{\zeta}=1.76\pm0.05.
\end{equation}

Using the linear approximates for $\xi(y)$ and $\zeta(y)$, we can calculate the semileptonic decay rates of processes $B_{s1}(B^{*}_{s2})\rightarrow D_{s}(D^{*}_{s})\ell\overline{\nu}$ and $B_{s1}(B^{*}_{s2})\rightarrow D_{s0}(D'_{s1})\ell\overline{\nu}$. The maximal values of $y$ for these semileptonic processes are given in Table \ref{table1}.
\begin{table}[h]
\caption{The maximal value of $y$ for each process: $y_{max}=(1+r_{i}^{2})/2r_{i}$ ($i=1, 2, \cdots, 8$).}
\begin{center}
\begin{tabular}{ccccccccc}
\hline \hline
 & & $D_{s}\ell\overline{\nu}$ & & $D^{*}_{s}\ell\overline{\nu}$ & & $D_{s0}\ell\overline{\nu}$ & & $D'_{s1}\ell\overline{\nu}$\\
\hline

$B_{s1}$  & & 1.64951 & & 1.56105 & & 1.45633 & & 1.396 \\

$B^{*}_{s2}$ & & 1.65183 & & 1.56317 & & 1.4582 & & 1.39772 \\

\hline \hline
\end{tabular}
\end{center}\label{table1}
\end{table}
By integrating the differential decay rates over the kinematic region $1 \leq y \leq y_{max}$, we get the decay widths of these semileptonic decay modes. Although the widths of $B_{s1}$ and $B^{*}_{s2}$ have not yet been measured experimentally, they were estimated early in Ref. \cite{OPAL95} to be around 1 MeV. Theoretically, their strong decays were investigated in Ref. \cite{LCL09}. As we know, the main decay modes of these excited $B_{s}$ mesons are strong decays. Therefore we can approximately take the strong decay widths as the total widths for an estimation of order of the branching ratios of these processes. In fact, the two-body strong decay widths of $B_{s1}$ and $B^{*}_{s2}$ are computed to be $98\mbox{keV}$ and $5\mbox{MeV}$ in Ref. \cite{LCL09}. Using these widths, we estimate the order of the branching ratios of the semileptonic decays (see Table \ref{table2}). Results of constituent quark meson (CQM) model in Ref. \cite{LCLZ09} are also shown there.
\begin{table}[h]
\caption{Predictions for the decay widths and branching ratios of $B_{s1}(B^{*}_{s2})\rightarrow D_{s}(D^{*}_{s})\ell\overline{\nu}$ and $B_{s1}(B^{*}_{s2})\rightarrow D_{s0}(D'_{s1})\ell\overline{\nu}$}
\begin{center}
\begin{tabular}{ccccccccc}
\hline \hline
Decay mode & & Decay width $\Gamma$ (GeV) & & BR & & one Result of Ref.\cite{LCLZ09} & & BR of Ref.\cite{LCLZ09}\\
\hline

$B_{s1}\rightarrow D_{s}\ell\overline{\nu}$ & & $(2.6\pm0.6)\times10^{-15}$ & & $\sim10^{-11}$ & & $2.1\times10^{-15}$ & & $\sim10^{-11}$ \\

$B_{s1}\rightarrow D^{*}_{s}\ell\overline{\nu}$ & & $(6.7\pm1.5)\times10^{-15}$ & & $\sim10^{-11}$ & & $4.9\times10^{-15}$ & & $\sim10^{-11}$ \\

$B^{*}_{s2}\rightarrow D_{s}\ell\overline{\nu}$ & & $(2.7\pm0.6)\times10^{-15}$ & & $\sim10^{-12}$ & & $2.1\times10^{-15}$ & & $\sim10^{-13}$ \\

$B^{*}_{s2}\rightarrow D^{*}_{s}\ell\overline{\nu}$ & & $(6.8\pm1.5)\times10^{-15}$ & & $\sim10^{-12}$ & & $5.0\times10^{-15}$ & & $\sim10^{-13}$ \\

$B_{s1}\rightarrow D_{s0}\ell\overline{\nu}$ & & $(1.0\pm0.3)\times10^{-18}$ & & $\sim10^{-14}$ & & $8.7\times10^{-20}$ & & $\sim10^{-16}$ \\

$B_{s1}\rightarrow D'_{s1}\ell\overline{\nu}$ & & $(1.9\pm0.5)\times10^{-18}$ & & $\sim10^{-14}$ & & $1.0\times10^{-19}$ & & $\sim10^{-16}$ \\

$B^{*}_{s2}\rightarrow D_{s0}\ell\overline{\nu}$ & & $(6.5\pm1.5)\times10^{-19}$ & & $\sim10^{-16}$ & & $5.6\times10^{-20}$ & & $\sim10^{-18}$ \\

$B^{*}_{s2}\rightarrow D'_{s1}\ell\overline{\nu}$ & & $(2.2\pm0.5)\times10^{-18}$ & & $\sim10^{-15}$ & & $1.2\times10^{-19}$ & & $\sim10^{-17}$ \\

\hline \hline
\end{tabular}
\end{center}\label{table2}
\end{table}

As we can see from Table \ref{table2}, the decay widths and branching ratios of  $B_{s1}(B^{*}_{s2})\rightarrow D_{s}(D^{*}_{s})\ell\overline{\nu}$ quite agree with the results of Ref. \cite{LCLZ09} while there are obvious deviations in $B_{s1}(B^{*}_{s2})\rightarrow D_{s0}(D'_{s1})\ell\overline{\nu}$. Our results are about one order higher than theirs in the latter cases. After a simple derivation, the formulas of differential decay widths in the present paper and Ref. \cite{LCLZ09} are exactly the same ((14)-(21) in Sec. \ref{sec2} and (19) from Ref. \cite{LCLZ09}). So the reason for the differences must be that the Isgur-Wise function $\zeta(y)$ estimated in the present paper and the corresponding  $\zeta(\omega)$ in Ref. \cite{LCLZ09} are different from each other. In fact, the Isgur-Wise function $\xi(y)$ here roughly agrees with $\xi(\omega)$ in Ref. \cite{LCLZ09} while $\zeta(y)$ is quite different from $\zeta(\omega)$ there. As mentioned above, the Isgur-Wise functions are estimated through QCD sum rules in this paper. The QCD sum rule method is a model independent method with its basis on first principle. It has been widely used for studying properties and decays of heavy-light mesons and proved to be a reliable method. In Ref. \cite{LCLZ09}, the authors calculated the
Isgur-Wise functions from CQM model which is an intermediate approach, not as rigorous and general as that of the effective meson Lagrangian, but
allows for a smaller number of input parameters \cite{Dea98}. In addition to that, the nonperturbative effect of QCD is systematically taken into
account in the QCD sum rule method while in CQM model, it is just treated as a suppression of large light quark momentum in terms of damping factors or cutoffs in the loop momentum integral \cite{Dea98}. To this end, it is reasonable to say that our results, which come from first principle calculations, are more reliable than model dependent calculations. Note that both calculations in this paper and Ref. [22] are confined at the leading order of heavy quark expansion, an account of corrections of higher order may improve the results.

Following the same way in which we estimate $\xi(y)$ and $\zeta(y)$ we can also calculate the three-point sum rules (\ref{rule3}) and (\ref{rule4}) for $\chi(y)$ and $\kappa(y)$ numerically. The decay constants and bounding energies that appear in (\ref{rule3}) and (\ref{rule4}) as input parameters are $f_{-,1/2}$, $f_{+,1/2}$, $\bar{\Lambda}_{-,1/2}$ and $\bar{\Lambda}_{+,1/2}$. The two-point sum rules we need to remove them are (\ref{conrule2}) and (\ref{conrule3}). The regions of the continuum thresholds are $2.3\mbox{GeV}<\omega_{c3}<2.7\mbox{GeV}$ and $2.7\mbox{GeV}<\omega_{c4}<3.1\mbox{GeV}$ for the three-point sum rules for $\chi(y)$ and $\kappa(y)$, respectively. The results are showed in Fig. 3 and Fig. 4 (In these figures, we fix $\omega_{0}=2.2\mbox{GeV}$, $\omega_{1}=2.9\mbox{GeV}$, and $\omega_{2}=3.0\mbox{GeV}$). Here we note that the Isgur-Wise function $\kappa(y)$ which describes the transition between heavy mesons with the same quantum numbers ($B_{s}(0^{+})\rightarrow D_{s}(0^{+})$) is approximately normalized as $\kappa(y=1)=1$, which is implied by the heavy quark symmetry \cite{Neu94}.
\begin{figure}
\begin{center}
\begin{tabular}{ccc}
\begin{minipage}{7cm} \epsfxsize=7cm
\centerline{\epsffile{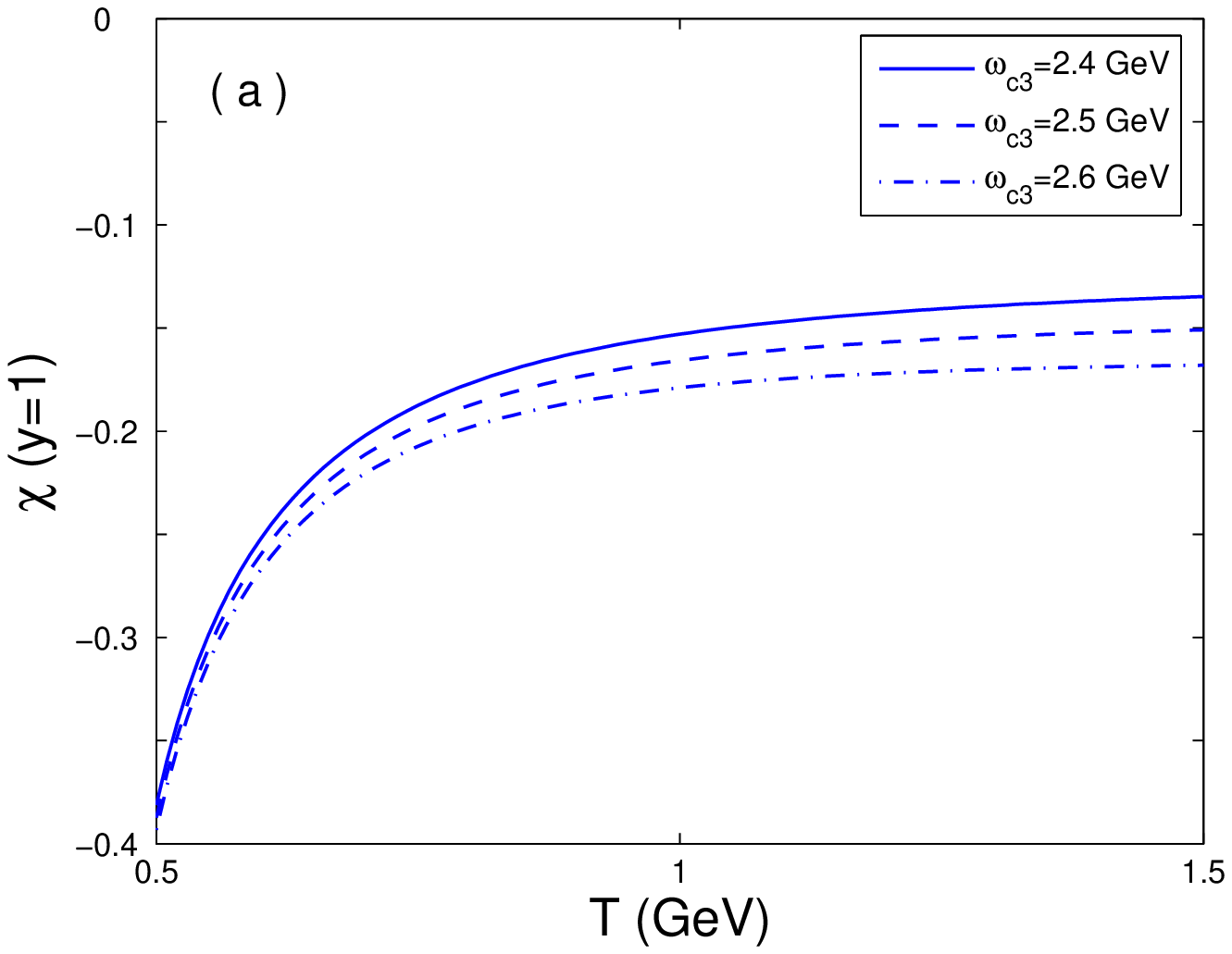}}
\end{minipage}& &
\begin{minipage}{7cm} \epsfxsize=7cm
\centerline{\epsffile{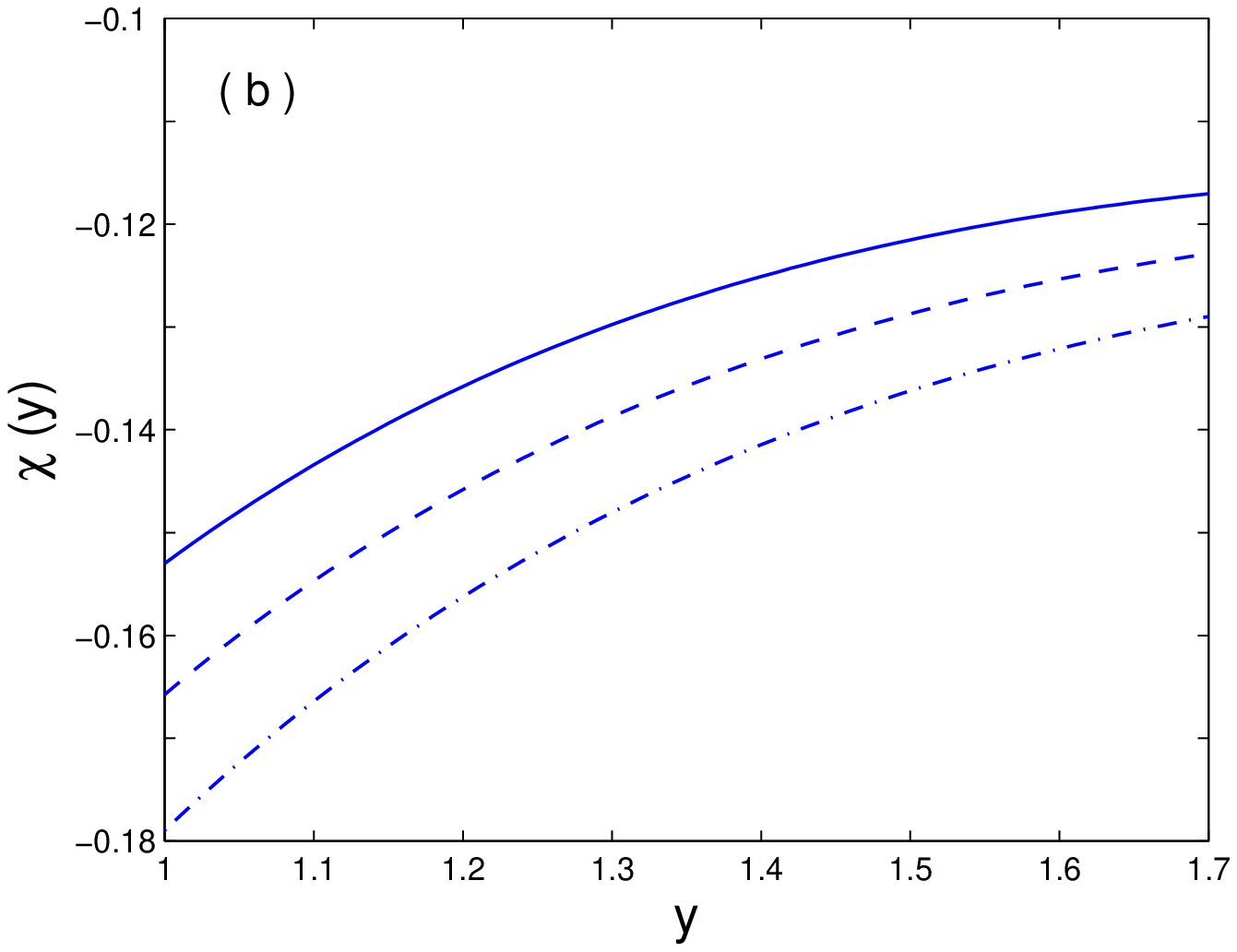}}
\end{minipage}
\end{tabular}
\caption{(a) {\it Dependence of  $\chi(y)$  on Borel parameter $T$ at $y=1$.} (b) {\it Prediction for the Isgur-Wise functions $\chi(y)$ at} $T=1 \mbox{GeV}$.}
\end{center}\label{fig3}
\end{figure}
\begin{figure}
\begin{center}
\begin{tabular}{ccc}
\begin{minipage}{7cm} \epsfxsize=7cm
\centerline{\epsffile{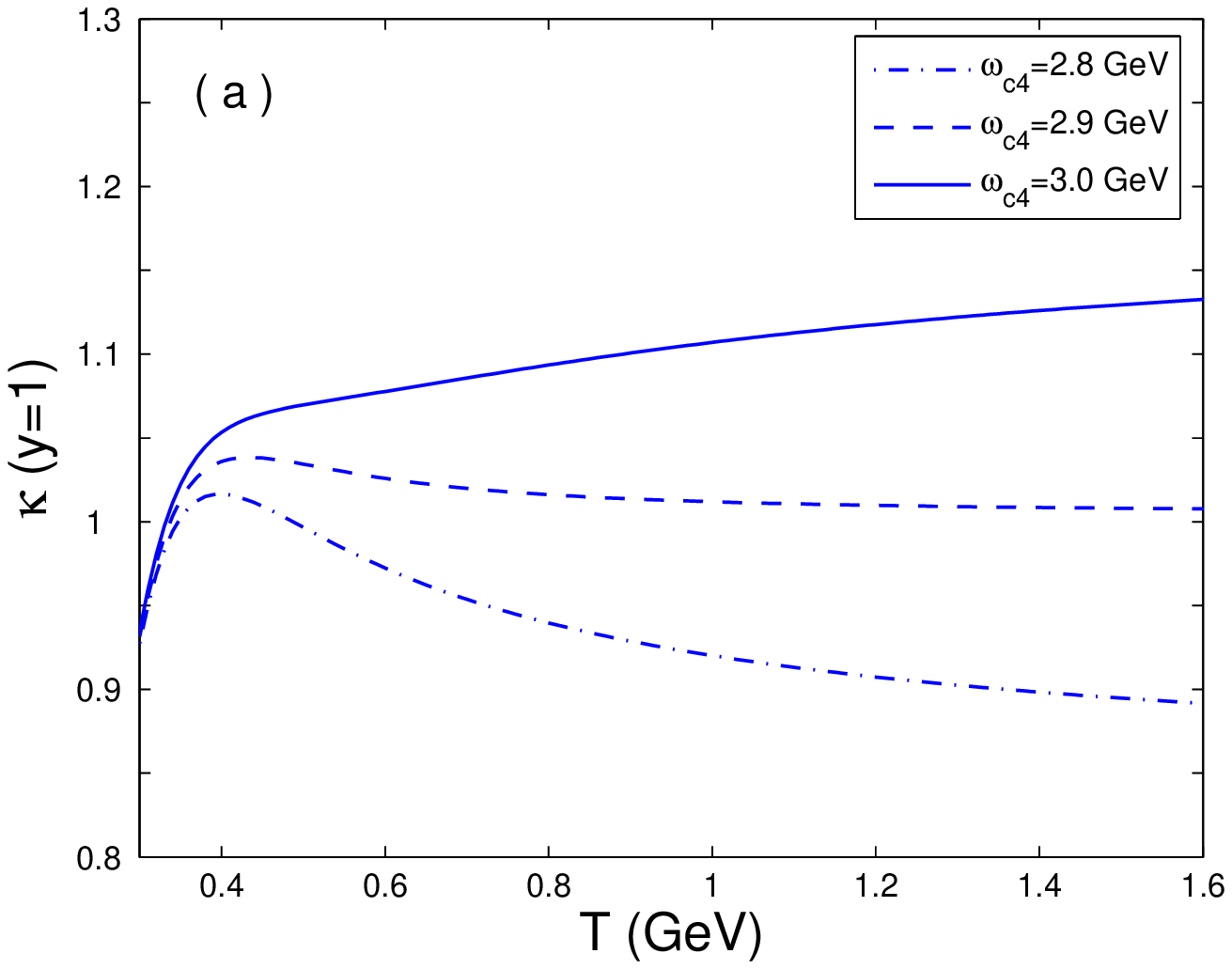}}
\end{minipage}& &
\begin{minipage}{7cm} \epsfxsize=7cm
\centerline{\epsffile{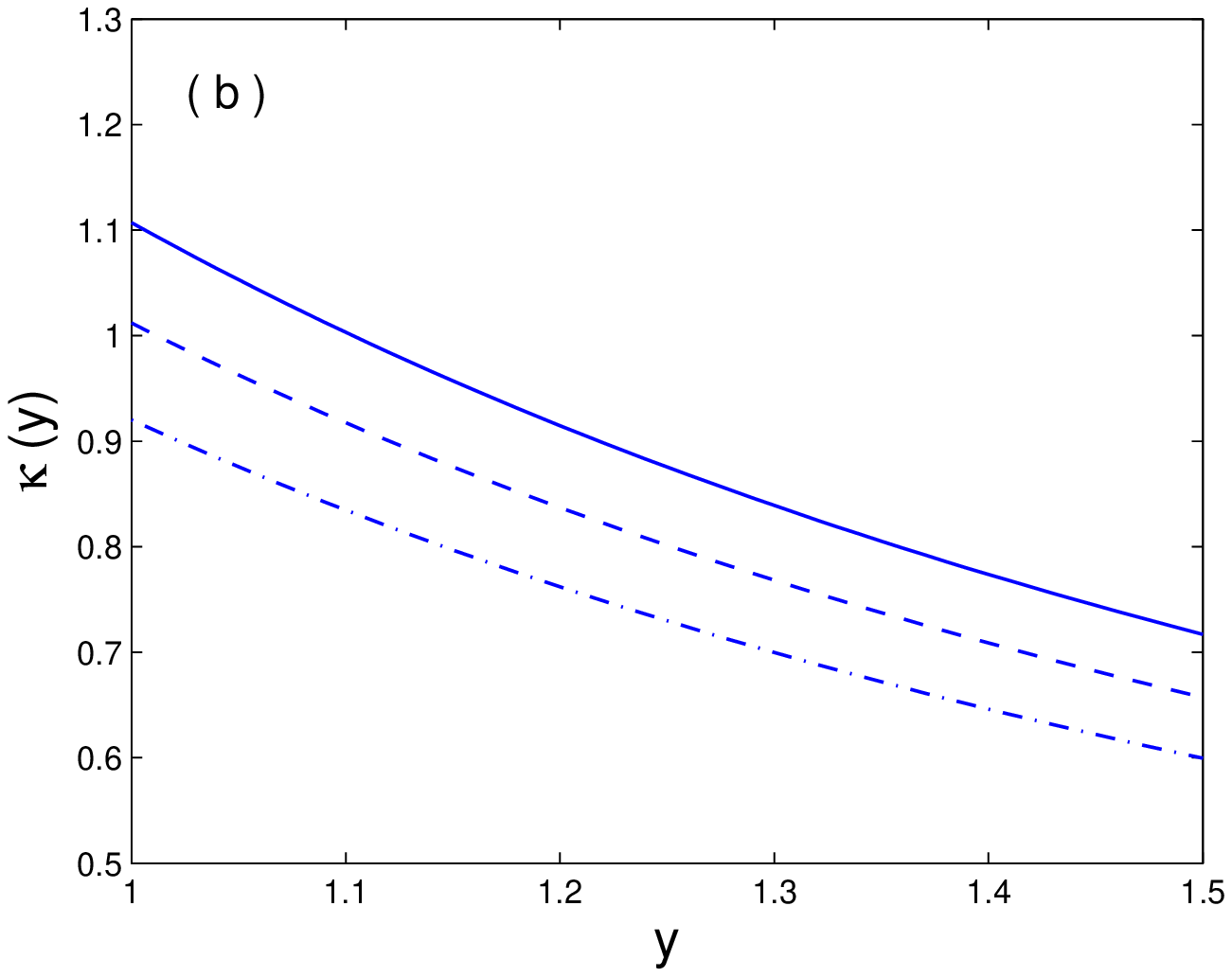}}
\end{minipage}
\end{tabular}
\caption{(a) {\it Dependence of  $\kappa(y)$  on Borel parameter $T$ at $y=1$.} (b) {\it Prediction for the Isgur-Wise functions $\kappa(y)$ at} $T=1 \mbox{GeV}$.}
\end{center}\label{fig4}
\end{figure}

The resulting curves for $\chi(y)$ and $\kappa(y)$ can be similarly parametrized by the linear approximation
\begin{equation}\label{linear3}
\chi(y)=\chi(1)[1-\rho^{2}_{\chi}(y-1)],\text{ }
\chi(1)=-0.16\pm0.03,\text{ }\rho^{2}_{\chi}=0.38\pm0.05;
\end{equation}
\begin{equation}\label{linear4}
\kappa(y)=\kappa(1)[1-\rho^{2}_{\kappa}(y-1)],\text{ }
\kappa(1)=0.99\pm0.09,\text{ }\rho^{2}_{\kappa}=0.7\pm0.1.
\end{equation}
Here we would like to give some remarks on the uncertainties of the sum rules (\ref{rule1}), (\ref{rule2}), (\ref{rule3}), and (\ref{rule4}). For the lack of information about higher resonances and continual state, the choices of the thresholds lead to the dominant uncertainties in the predictions for the Isgur-Wise functions. The input nonperturbative parameters, namely the values of the vacuum condensates, will also lead to errors to the form factors. In our calculation, only the errors due to the choices of sum rule windows are considered.

The maximal value of $y$ for each process of semileptonic $B_{s0}$ and $B'_{s1}$ decays is showed in Table \ref{table3}.
\begin{table}[h]
\caption{The maximal value of $y$ for each process: $y_{max}=(1+r_{i}^{2})/2r_{i}$ ($i=9, 10, \cdots, 16$).}
\begin{center}
\begin{tabular}{ccccccccc}
\hline \hline
 & & $D_{s}\ell\overline{\nu}$ & & $D^{*}_{s}\ell\overline{\nu}$ & & $D_{s0}\ell\overline{\nu}$ & & $D'_{s1}\ell\overline{\nu}$\\
\hline

$B_{s0}$ & & 1.62451 & & 1.53821 & & 1.43617 & & 1.37746 \\

$B'_{s1}$ & & 1.63504 & & 1.54783 & & 1.44466 & & 1.38526 \\

\hline \hline
\end{tabular}
\end{center}\label{table3}
\end{table}
Using these maximal values and the forms of linear approximations for $\chi(y)$ and $\kappa(y)$, one can compute the decay widths of $B_{s0}(B'_{s1})\rightarrow D_{s}(D^{*}_{s})\ell\overline{\nu}$ and $B_{s0}(B'_{s1})\rightarrow D_{s0}(D'_{s1})\ell\overline{\nu}$. Considering that the main decay modes of $B_{s0}$ and $B'_{s1}$ are isospin violating decays and radiative decays, they are supposed to have widths of about 100 keV \cite{BEH03}. One can then roughly estimate the branching ratios of these decays. All the results are presented in Table \ref{table4}.
\begin{table}[h]
\caption{Predictions for the decay widths and branching ratios of $B_{s0}(B'_{s1})\rightarrow D_{s}(D^{*}_{s})\ell\overline{\nu}$ and $B_{s0}(B'_{s1})\rightarrow D_{s0}(D'_{s1})\ell\overline{\nu}$}
\begin{center}
\begin{tabular}{ccccccccc}
\hline \hline
Decay mode & & Decay width $\Gamma$ (GeV) & & BR & & one Result of Ref.\cite{LCLZ09} & & BR of Ref.\cite{LCLZ09}\\
\hline

$B_{s0}\rightarrow D_{s}\ell\overline{\nu}$ & & $(1.3\pm0.5)\times10^{-16}$ & & $\sim10^{-12}$ & & $2.5\times10^{-14}$ & & $10^{-9}\sim10^{-10}$ \\

$B_{s0}\rightarrow D^{*}_{s}\ell\overline{\nu}$ & & $(1.5\pm0.5)\times10^{-16}$ & & $\sim10^{-12}$ & & $2.5\times10^{-15}$ & & $10^{-9}\sim10^{-10}$ \\

$B'_{s1}\rightarrow D_{s}\ell\overline{\nu}$ & & $(6.6\pm2.3)\times10^{-17}$ & & $\sim10^{-12}$ & & $1.2\times10^{-14}$ & & $10^{-9}\sim10^{-10}$ \\

$B'_{s1}\rightarrow D^{*}_{s}\ell\overline{\nu}$ & & $(2.1\pm0.7)\times10^{-16}$ & & $\sim10^{-12}$ & & $3.8\times10^{-14}$ & & $10^{-9}\sim10^{-10}$ \\

$B_{s0}\rightarrow D_{s0}\ell\overline{\nu}$ & & $(1.2\pm0.3)\times10^{-14}$ & & $\sim10^{-10}$ & & $1.6\times10^{-15}$ & & $10^{-10}\sim10^{-11}$ \\

$B_{s0}\rightarrow D'_{s1}\ell\overline{\nu}$ & & $(3.4\pm0.6)\times10^{-14}$ & & $\sim10^{-10}$ & & $1.5\times10^{-15}$ & & $10^{-9}\sim10^{-10}$ \\

$B'_{s1}\rightarrow D_{s0}\ell\overline{\nu}$ & & $(4.0\pm0.7)\times10^{-14}$ & & $\sim10^{-10}$ & & $4.9\times10^{-16}$ & & $10^{-10}\sim10^{-11}$ \\

$B'_{s1}\rightarrow D'_{s1}\ell\overline{\nu}$ & & $(1.1\pm0.2)\times10^{-13}$ & & $\sim10^{-9}$ & & $2.1\times10^{-15}$ & & $10^{-9}\sim10^{-10}$ \\

\hline \hline
\end{tabular}
\end{center}\label{table4}
\end{table}

We would like to address that $B_{s0}$ and $B'_{s1}$ are still missing in experiments. Their masses were theoretically estimated through various methods which gave quite different values \cite{BEH03,VVF08,EHQ93}. The branching ratios given in Table \ref{table4} are calculated on the assumption that $B_{s0}$ and $B'_{s1}$ lie below the thresholds of $B^{*}K$ and $BK$. If this assumption is not true, $B_{s0}$ and $B'_{s1}$ can decay through these modes and they both own a width of hundreds of MeV. Then the branching ratios of the semileptonic decays in Table \ref{table4} should be 3 orders of magnitude lower. The present precision of the experimental measurement of the branching ratio of the $B_{s}$ mesons has reached up to $10^{-7} \sim 10^{-8}$ \cite{PDG08}. As we can see from Table \ref{table2} and Table \ref{table4}, our numerical results indicate that the decay widths of these semileptonic modes are tiny and therefore exclude the possibility of finding them in experiments. The decays $B_{s0}(B'_{s1})\rightarrow D_{s0}(D'_{s1})\ell\overline{\nu}$ may be expected to be measured in the future LHCb experiment. If the experimental results agree with our prediction, it can be supportive to the ordinary $c\bar{s}$ meson explanation for $D_{sJ}(2317)$ and $D_{sJ}(2460)$.

In summary, we have performed a study of the semileptonic decays of $B^{**}_{s}$ mesons into low lying charmed-strange mesons within the framework of HQET. Two states out of the four $P$-wave excited $B_{s}$ mesons, $B_{s1}$ and $B^{*}_{s2}$, have been recently measured by CDF and D0 Collaborations while the other two, $B_{s0}$ and $B'_{s1}$, are still missing. We employ QCD sum rules to estimate the leading-order universal form factors describing the transitions of these $B^{**}_{s}$ states into low lying $D_{s}$ mesons, including $D_{s}(1968)$, $D^{*}_{s}(2112)$, $D_{sJ}(2317)$, and $D_{sJ}(2460)$. The predicted branching ratios of these processes are prohibitively tiny. It is worth noting that the Isgur-Wise function $\kappa(y)$ which parameterizes the hadronic matrix element of the weak current between the $S$ doublet of $\bar{b}s$ system and the same doublet of $\bar{c}s$ system approximately satisfies the normalization condition, $\kappa(y=1)=1$, which is implied by the heavy quark flavor symmetry at the leading order of the heavy quark expansion. Although the branching ratios we predicted are very small, some of them are expected to be possibly observed in the forthcoming LHCb experiment. A measurement of these processes can provide some information on the structure of the $D_{sJ}(2317)$ and $D_{sJ}(2460)$ mesons.

\begin{acknowledgments}
This work was supported in part by the National Natural Science Foundation of China under Contract No. 10975184.
\end{acknowledgments}

\end{document}